\documentclass[%
 reprint,
superscriptaddress,
 amsmath,amssymb,
 aps,
prb,
floatfix,
]{revtex4-2}

\usepackage{enumitem}
\usepackage{graphicx}
\usepackage{dcolumn}
\usepackage{bm}
\usepackage[mathlines]{lineno}
\usepackage{appendix}
\usepackage{braket}
\usepackage[dvipsnames]{xcolor}
\usepackage{amsmath,amssymb,amsfonts} 
\usepackage{makecell}
\usepackage{float}
\usepackage{bbm}
\usepackage{nicefrac}
\usepackage{wasysym}
\usepackage[normalem]{ulem}
\usepackage{mathrsfs}
\usepackage{hyperref}
\hypersetup{
	colorlinks=true,
	linkcolor=blue,
	filecolor=blue,      
	urlcolor=blue,
	citecolor=blue,
}

\begin{document}

\preprint{APS/123-QED}

\title{Noise-Induced Resurrection of Dynamical Skin Effects in Quasiperiodic Non-Hermitian Systems}

\author{Wuping Yang}
\affiliation{School of Physics, Peking University, Beijing 100871, China}

\author{H. Huang}%
\email[Contact author: ]{huanghq07@pku.edu.cn}
\affiliation{School of Physics, Peking University, Beijing 100871, China}
\affiliation{Collaborative Innovation Center of Quantum Matter, Beijing 100084, China}
\affiliation{Center for High Energy Physics, Peking University, Beijing 100871, China}

\date{\today}
             
\begin{abstract}
The non-Hermitian skin effect (NHSE) refers to the accumulation of an extensive number of eigenstates at system boundaries under open boundary conditions (OBCs). As a dynamical consequence, wave packets in such systems drift and ultimately accumulate at a boundary, giving rise to the dynamical skin effect (DSE). While strong quasiperiodic potentials are known to suppress the DSE by inducing localization, we show that the introduction of Ornstein-Uhlenbeck (OU) noise unexpectedly restores it. Using perturbative analysis, we demonstrate that noise effectively maps the non-Hermitian Schr\"{o}dinger dynamics onto a non-reciprocal master equation, whose complex spectrum develops a noise-induced point gap. This mechanism enables delocalization, reinstates directional transport, and revives the DSE even in regimes where the static NHSE is absent. Moreover, the relaxation dynamics exhibit a non-monotonic dependence on noise strength, reflecting a competition between noise-assisted delocalization and noise-induced decoherence. Our results uncover a noise-enabled mechanism for resurrecting the DSE and suggest a new route for controlling transport in quasiperiodic, open quantum systems.
\end{abstract}

\maketitle


\section{\label{sec:level1} INTRODUCTION}
Non-Hermitian (NH) Hamiltonians provide a powerful framework for describing open quantum systems, where environmental coupling plays a crucial role~\cite{RevModPhys.93.015005, doi:10.1080/00018732.2021.1876991,PhysRevLett.85.2478,PhysRevC.67.054322,Rotter_2009,PhysRevA.85.032111,Rotter_2015}. Such effective models have been widely realized in classical wave platforms, including photonic~\cite{RevModPhys.91.015006, feng2017Non, AosRnsurr2012Parity, doi:10.1126/science.1258479, 2015Spawning, doi:10.1126/science.aap9859, doi:10.1126/science.aar4005, Nasari:23,PhysRevLett.127.270602}, acoustic~\cite{PhysRevX.6.021007, doi:10.1126/science.abd8872, PhysRevLett.121.085702, PhysRevLett.127.034301, PhysRevLett.118.174301, PhysRevApplied.16.014012}, and electrical circuit systems~\cite{2020Generalized,0Observation,PhysRevB.107.085426,advs.202301128}.  
A hallmark of NH systems is the non-Hermitian skin effect (NHSE), in which a macroscopic number of eigenstates become exponentially localized at the boundaries under OBCs~\cite{PhysRevLett.121.086803}. The NHSE also manifests dynamically: wave packets evolving under OBCs inevitably drift toward the boundary, resulting in the dynamical skin effect (DSE)~\cite{PhysRevB.106.L241112,li2024observation}. Recent studies have uncovered additional DSE-driven dynamical phenomena, including self-healing~\cite{PhysRevLett.128.157601} and edge bursts~\cite{PhysRevLett.128.120401}.

Meanwhile, one-dimensional quasicrystals with aperiodic order provide an ideal platform for exploring anomalous transport phenomena across classical and quantum settings~\cite{PhysRevLett.120.160404,PhysRevLett.103.013901,PhysRevLett.109.106402,WOS:000505617400023,WOS:000470795000001}. Quasiperiodicity generates a variety of unconventional behaviors—critical energy spectra, fractal eigenstates, and localization transitions, among others~\cite{PhysRevLett.51.1198,PhysRevB.28.4272,PhysRevA.36.5349,PhysRevLett.66.1651,PhysRevLett.68.3826,PhysRevLett.73.3379,PhysRevLett.76.4372,PhysRevB.50.1420,PhysRevLett.81.3735}. In non-reciprocal NH systems, quasiperiodicity has been shown to qualitatively alter wave-packet dynamics: across the localization transition, the wave-packet velocity exhibits a discontinuous jump, separating directional transport in the DSE phase from dynamical localization in the localized phase~\cite{PhysRevB.103.054203}.  

In Hermitian systems, it is known that temporal noise can dynamically restore diffusion in localized phases, effectively counteracting localization over long timescales~\cite{PhysRevLett.119.046601}. This observation naturally raises an open and unexplored question:  
\emph{Can noise similarly induce delocalization in quasiperiodic non-Hermitian systems where localization has suppressed the DSE?}

In this work, we address this question by investigating the interplay of decoherence noise, quasiperiodic localization, and non-Hermiticity in a non-reciprocal quasicrystal. We demonstrate that Ornstein–Uhlenbeck (OU) noise can resurrect the DSE even deep inside the localized regime, thereby enabling a new form of noise-driven transport in open quantum systems. Using perturbative analysis, we show that noise effectively reduces the non-Hermitian Schr\"{o}dinger equation to a non-reciprocal master equation, providing a clear mechanism for the restoration of directional flow. Our framework additionally reveals distinct dynamical scaling behaviors of wave packets at short and long times. Additionally, we uncover a non-monotonic dependence of the effective transport coefficient on the noise amplitude: it initially increases and then decreases, reflecting an unconventional relaxation process arising from the competition among quasiperiodicity, non-Hermiticity, and noise.

\begin{figure*}[!htbp] 
    \centering
    \includegraphics[width=1\linewidth]{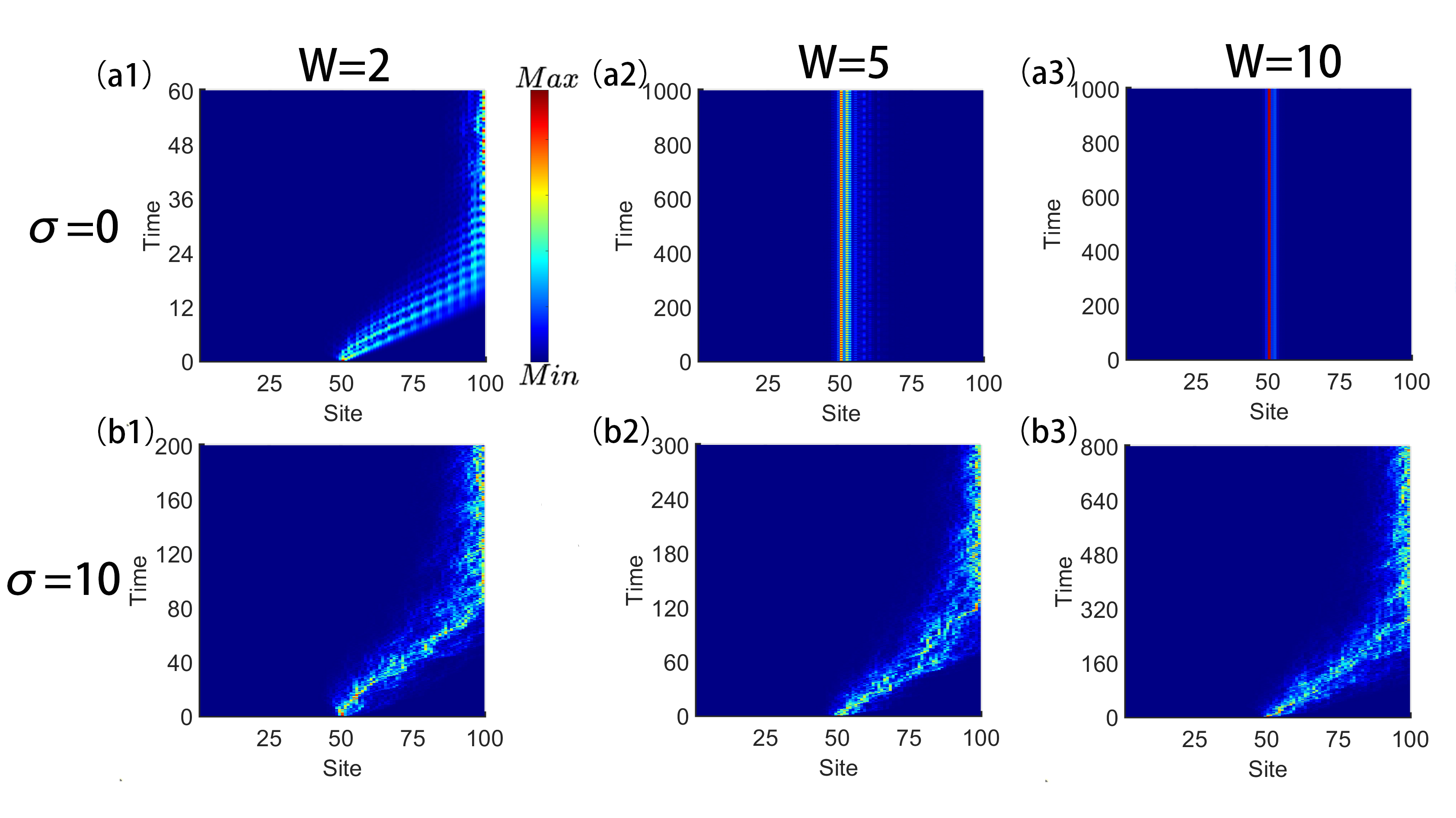}
    \caption{ \label{fig2} Time evolution of the wave function under different noise strengths $\sigma$ and quasiperiodic strengths $W$. System parameters are $J=\frac{3}{2} , \Delta=\frac{1}{2}$, and the system size is chosen as $L = 100$. The initial state is localized at the center of the one-dimensional chain.}
    \label{fig1}
\end{figure*}

\section{MODEL AND METHODS}

\subsection{\label{sec:level2}One-dimensional non-reciprocal quasicrystal model}
We consider a non-reciprocal Aubry-Andr\'e-Harper (AAH) model subject to a time-dependent noise generated by an Ornstein-Uhlenbeck (OU) process. The Hamiltonian is

\begin{equation}
\begin{split}
\hat{H} = &\sum_{j} \left[ (J + \Delta) \hat{c}_{j}^{\dagger} \hat{c}_{j-1} + (J - \Delta) \hat{c}_{j}^{\dagger} \hat{c}_{j+1} \right.\\
        &\left. + \left( W \cos(2 \pi \beta j) + \xi(j,t) \right) \hat{c}_{j}^{\dagger} \hat{c}_{j}\right],
\end{split}\label{Ham}
\end{equation}
where $c_{j}^{\dagger}$ ($c_{j}$) creates (annihilates) a fermion at site $j$. The asymmetric hopping amplitudes $J \pm \Delta$ break Hermiticity and generate non-reciprocity, while $W$ denotes the strength of the quasiperiodic potential. We fix $\beta$ to $\left( \sqrt{5}-1\right)/2$ to ensure incommensurability. 

The noise $\xi(j,t)$ is independently generated at each lattice site via an OU process with temporal correlations:
\begin{equation}
\langle \xi(j,t)\xi(j',t+\tau) \rangle =\frac{\sigma^{2}}{2\theta} e^{-\theta|\tau|}\delta_{jj'},
\end{equation}
where $\sigma$ is the noise strength and $\theta$ is the reversion rate (see Supplementary Material (SM) Sec. I for details ~\cite{fn}. Unless otherwise specified, we set $\theta=1$ in all numerical simulations.  

In the clean limit without quasiperiodic potential or noise (i.e., $W$=0, $\xi(l,t)=0$), the system exhibits the NHSE: for $\Delta > 0$ ($\Delta < 0$), all eigenstates accumulate at the right (left) boundary under OBCs. 
When a static quasiperiodic potential is included ($W\neq 0$, $\xi(l,t)=0$), previous works \cite{PhysRevB.103.054203,PhysRevB.104.024201} identified a localization-delocalization transition 
at the critical strength of the quasiperiodic potential
\begin{equation}
W_{c}=2\max \{\,|J+\Delta|,\; |J-\Delta|\,\}.
\end{equation}
For $W < W_{c}$, all eigenstates remain non-Hermitian skin modes (delocalized phase), whereas for $W > W_{c}$, all states become exponentially localized with a purely real energy spectrum (localized phase).

\section{NOISE RESURRECTS THE DYNAMIC SKIN EFFECT}

\subsection{Evolution of Wave Functions} 
Fig.~\ref{fig1} illustrates the time evolution of wave functions for varying noise strengths $\sigma$ and quasiperiodic potentials $W$. Throughout this subsection, system parameters are fixed at $J = 3/2$, $\Delta = 1/2$, and chain length $L = 100$. The initial state is a wave packet localized at the center of the chain. Each column displays results for a different quasiperiodic strength: (a) $W = 2$ (delocalized phase), (b) $W = 5$ (localized phase), and (c) $W = 10$ (deeply localized phase). Each row corresponds to a different noise strength: $\sigma = 0$ and $\sigma = 10$. 

For $W = 2 < W_c$, the asymmetric hopping induces inherently unidirectional transport. In the absence of noise ($\sigma = 0$), the wave packet expands according to a half light cone~\cite{WOS:000360235700001, Jin_2017} and accumulates at the right boundary, as shown in Fig.~\ref{fig1}(a1). Introducing OU noise ($\sigma = 10$) enhances fluctuations in the propagation path but leaves the overall directed motion and boundary accumulation intact, as shown in Fig.~\ref{fig1}(b1).

In the localized regime ($W = 5, 10$) without noise, the quasiperiodic potential suppresses the NHSE: all eigenstates are exponentially localized, and the wave packet remains confined around its initial position with no drift or expansion [Fig.~\ref{fig1}(a2),(a3)]. Remarkably, adding OU noise ($\sigma = 10$) restores directional propagating and boundary accumulation. As shown in Fig.~\ref{fig1} (b2),(b3), the wave packet delocalizes, propagates toward the right, and eventually accumulates at the boundary, demonstrating a noise-induced revival of the DSE.

This mechanism can be understood intuitively: the quasiperiodic potential traps particles in deep potential minima, suppressing transport. A time-dependent noisy potential intermittently lowers or reshapes these energy barriers, allowing particles to escape from localized wells and restoring the collective directional drift characteristic of the DSE.

\begin{figure*}[!htbp] 
    \centering
    \includegraphics[width=1\linewidth]{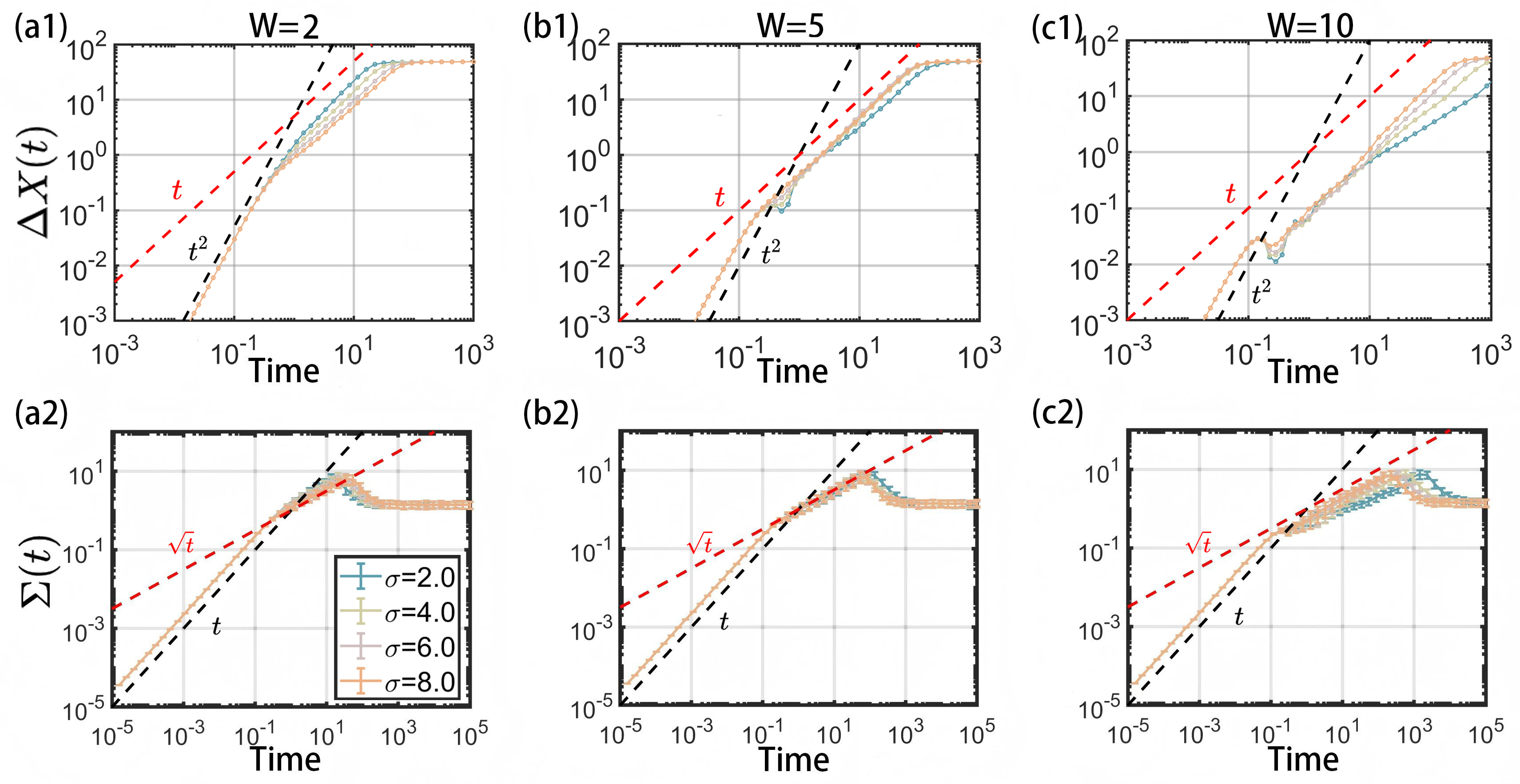}
    \caption{ \label{fig3} Time evolution of the increment of mean position $\Delta X\left( t\right) \equiv X\left( t\right) -X\left( 0\right)$ and spread moment $\Sigma(t)$ under different noise strengths $\sigma$ and quasiperiodic strengths $W$. The system parameters are selected as $J=\frac{3}{2} , \Delta=\frac{1}{2}$, and the system size is chosen as $L = 100$. The initial state is selected as the state localized at the center of the one-dimensional chain. The ensemble average is calculated over 100 random samples for each data point. In (a1), (b1), and (c1), the black and red dashed lines represent $\Delta X\left( t\right)\sim t^{2}$ and $\Delta X\left( t\right)\sim t$.  In (a2) (b2) (c2), the black dashed line represents the function $\Sigma(t)\sim t$, while the red dashed line represents $\Sigma(t) \sim \sqrt {t}$. }
    \label{fig2}
\end{figure*}

Furthermore, while the DSE is the dynamical signature of the NHSE in clean systems, our results demonstrate that the two are not mutually dependent. A noisy time-dependent perturbation can resurrect the DSE even in regimes where all eigenstates of the static Hamiltonian are localized and the NHSE is absent. This establishes the DSE as an intrinsically dynamical and noise-resilient phenomenon that can persist beyond the conventional static NHSE framework.

\subsection{Mean position and spread moment}
To further quantitatively characterize the noise-driven dynamics in the NH system, we define the mean position
\begin{equation}\label{3}
X(t)=\sum_{j} j|A_{j}(t)|^{2}
\end{equation}
and spread moment\cite{PhysRevLett.119.046601}
\begin{equation}\label{4}
\Sigma(t)=\left[\sum_{j} j^{2}|A_{j}(t)|^{2}-\left(\sum_{j} j|A_{j}(t)|\right)^{2}\right]^{1/2},
\end{equation}
where $A_{j}(t)$ is the wave function amplitude at lattice site $j$ and time $t$.
The temporal evolution of $X(t)$ measures the directed motion of the wave packet, while $\Sigma(t)$ captures its spatial variance. In order to capture the temporal growth scaling behavior of $X(t)$ over time, it is beneficial to further introduce the increment of mean position $\Delta X\left( t\right) \equiv X\left( t\right) -X\left( 0\right)$. 
 The scaling of $\Sigma(t)$ distinguishes transport regimes: ballistic ($\Sigma(t)\sim t$) versus diffusive ($\Sigma(t)\sim\sqrt{t}$)~\cite{PhysRevLett.119.046601}.

Figs. \ref{fig2}(a1)-(c2) summarizes the behavior of $\Delta X\left( t\right)$ and $\Sigma(t)$ for varying noise strengths and quasiperiodic potentials. Each column corresponds to $W = 2, 5,$ and $10$, respectively, and each panel compares different noise strengths $\sigma = 2, 4, 6, 8$.
{At short times, $\Delta X\left( t\right)$ grows quadratically} while $\Sigma(t)$ increases linearly, indicating ballistic propagation. The curves for different $\sigma$ collapse perfectly in this regime, showing that the initial expansion is insensitive to both the quasiperiodic potential and the noise. 
{Following this ballistic regime, the system transitions to a diffusive stage where 
$\Sigma(t)\propto{t}^{1/2}$ and $\Delta X\left( t\right)\propto{t}$} \footnote{Notably, in the regime of strong quasiperiodicity with relatively weak noise (e.g., $W=10$ and $\sigma=2$), $\Sigma(t)$ exhibits subdiffusive scaling behavior. Simultaneously, $\Delta X(t)$ is characterized by a scaling exponent slightly smaller than that of the standard linear drift regime, i.e., $\Delta X(t) \sim t^{\alpha}$ with $\alpha \lesssim 1$.} . The eventual saturation of 
$\Sigma(t)$ and $\Delta X\left( t\right)$ is a finite-size effect arising when the wave packet reaches the system boundary.

We can define the relaxation time $\tau_{\rm relax}$ as the characteristic timescale of this saturation for a fixed system size.
Remarkably, the relaxation time $\tau_{\text{relax}}$ exhibits a non-monotonic dependence on $\sigma$ that strongly depends on $W$. For weak quasiperiodicity ($W=2$), $\tau_{\text{relax}}$ increases with $\sigma$ (compare $\sigma=2$ and $\sigma=8$ in Fig.~\ref{fig2}(a1)), suggesting that noise disrupts the coherent directional bias provided by non-reciprocal hopping, thereby slowing the DSE. In contrast, for strong quasiperiodicity ($W=10$), $\tau_{\text{relax}}$ decreases with $\sigma$ (Fig.~\ref{fig2}(c1)), reflecting noise-enhanced delocalization that restores directional motion. At intermediate potential strength ($W=5$), $\tau_{\text{relax}}$ is nearly insensitive to $\sigma$, consistent with a competition between noise-induced hopping and noise-induced decoherence (A detailed numerical analysis of $\tau_{\text{relax}}$  can be found in Section V of the SM.~\cite{fn}). This reversal in behavior highlights a subtle interplay among quasiperiodicity, non-Hermiticity, and temporally correlated noise: weak quasiperiodicity suppresses directional motion while strong quasiperiodicity requires noise to re-enable it.

\section{Perturbative Analysis}

\subsection{Perturbative treatment in the strong-noise regime}
To gain analytic insight into the anomalous transport behaviors observed numerically, we develop a perturbative treatment valid in the strong-noise regime ($\sigma\gg J, \Delta$). The single-particle Schr\"{o}dinger equation in first quantization reads
\begin{equation}\label{7}
\begin{split}
i \partial_{t} A_{j}(t) &= (J+\Delta) A_{j+1}(t)  + (J-\Delta) A_{j-1}(t) \\
&\quad+ \xi(j,t) A_{j}(t)  + W \cos(2\pi\beta j) A_{j}(t),
\end{split}
\end{equation}

We first consider the zero-hopping limit ($J=\Delta=0$), in which lattice sites become completely decoupled. The evolution of the amplitude at site $j$ is then
\begin{equation}\label{8}
A_{j}^{(0)}(t) = A_{j}^{(0)}(0) e^{-i\varepsilon_j t -i \int_0^t \xi_j(\tau) d\tau},
\end{equation}
with on-site energy $\varepsilon_j = W \cos(2\pi\beta j)$. 

We expand the full amplitude as $A_j(t)$ as $A_j(t) = A_j^{(0)}(t) + A_j^{(1)}(t)$ and substitute into Eq.~\eqref{7}. Retaining only terms linear in $J$ and $\Delta$ yields the first-order correction
\begin{equation}\label{9}
\begin{aligned}
A_{j}^{(1)}= & e^{-i \phi_{j}(t)} \int_{0}^{t}\left(\frac{J}{i}\right) e^{i \phi_{j}(\tau)}\left[A_{j+1}^{(0)}(\tau)+A_{j-1}^{(0)}(\tau)\right] d \tau \\
& +e^{-i \phi_{j}(t)} \int_{0}^{t}\left(\frac{\Delta}{i}\right) e^{i \phi_{j}(\tau)}\left[A_{j+1}^{(0)}(\tau)-A_{j-1}^{(0)}(\tau)\right] d \tau,
\end{aligned}
\end{equation}
where $\phi_j(t)\equiv \varepsilon_j t +\int_0^t \xi(j,s)ds$.

To extract the corresponding transport behavior, we examine the probability evolution equation:
\begin{equation}\label{pro}
\begin{aligned}
\frac{d}{d t} P_{j}= & 2 J \operatorname{Im}\left(A_{j}^{*} A_{j+1}+A_{j}^{*} A_{j-1}\right) \\
& +2 \Delta \operatorname{Im}\left(A_{j}^{*} A_{j+1}-A_{j}^{*} A_{j-1}\right), 
\end{aligned}
\end{equation}
where $P_{j}\equiv \left| A_{j}\right| ^{2}$ represents the probability of particles appearing at site $j$.
Substituting Eqs.~\eqref{8} and \eqref{9} into Eq.~\eqref{pro} and retaining terms up to the first order yields the noise master equation (full derivation in the  SM.~\cite{fn}):
\begin{equation}\label{11}
\begin{aligned}
\frac{d}{d t}\left\langle P_{j}\right\rangle = & 2\left\{\operatorname{Re}\left(Q_{j, j+1}+Q_{j, j-1}\right)(\Delta+J)(\Delta-J)\left\langle P_{j}\right\rangle\right. \\
& + \operatorname{Re}\left(Q_{j, j+1}\right)(\Delta+J)^{2}\left\langle P_{j+1}\right\rangle \\
& \left. + \operatorname{Re}\left(Q_{j, j-1}\right)(\Delta-J)^{2}\left\langle P_{j-1}\right\rangle\right\},
\end{aligned}
\end{equation}
where the noise-dependent kernel is given by:
\begin{equation}\label{10}
\begin{aligned}
\operatorname{Re} Q_{j, j+1} 
& =\int_{0}^{t} \cos \left[\left(\varepsilon_{j}-\varepsilon_{j+1}\right)(t-\tau)\right]\left|C_{\phi}(t-\tau)\right|^{2} d \tau, \\
\end{aligned}
\end{equation}
with $C_{\phi}(t-\tau) \equiv\langle e^{i \int_{\tau}^{t} \xi(j,s) d s}\rangle$, and $\langle \ldots \rangle$ denoting noise averaging.
In fact, the kernel $\operatorname{Re} Q_{j, j+1}$ is a key transport quantifier, as we show below. For a more detailed derivation and comprehensive discussion of $\operatorname{Re} Q_{j, j+1}$, please refer to the SM.~\cite{fn}.

\subsection{Estimation of ${\operatorname{Re} Q_{j, j+1}}$}
The central quantity governing noise-assisted transport is the kernel $\operatorname{Re} Q_{j, j+1}$.
Although Eq. (\ref{10}) cannot be solved exactly in closed form for general $t$, reliable approximations can be obtained in the short-time and long-time limits. For $t\rightarrow 0$ in the short-time limit,  expanding the integrand yields
\begin{equation}\label{13}
\begin{aligned}
{\operatorname{Re} Q_{j, j+1}} &\approx  t-\frac{1}{6}\left[4W^2\sin^2(\pi\beta(2j+1))\sin^2(\pi\beta) +
\frac{\sigma^2}{\theta}\right]t^3 .
\end{aligned}
\end{equation}
The influence of both quasiperiodic strength $W$ and the noise intensity $\sigma$ enter $\operatorname{Re} Q_{j, j+1}$ only at order $t^3$. Consequently, the early-time dynamics are insensitive to either type of disorder. This explains the universal short-time collapse of the $\Delta X(t)$ and $\Sigma(t)$ observed numerically in Fig.~\ref{fig2}.

For $t\rightarrow \infty$ in the long-time limit, the kernel $\operatorname{Re} Q_{j, j+1}$ saturates to
\begin{equation}\label{12}
\begin{aligned}
\operatorname{Re} Q_{j, j+1} \approx \frac{\sigma^2\theta^2}{\sigma^4+4W^2\theta^4\sin^2(\pi\beta(2j+1))\sin^2(\pi\beta)}.
\end{aligned}
\end{equation}
Noticeably, $\operatorname{Re} Q_{j, j+1}$ in this regime exhibits explicit spatial dependence through the lattice site index $j$. 
To describe global transport properties, we therefore introduce the spatial average $\overline{\operatorname{Re} Q_{j, j+1}}$, 
which provides a unified quantification of transport efficiency across the entire system. 
This averaged quantity allows Eq. (\ref{11}) to be approximated as 
\begin{equation}\label{discrete}
\begin{split}
\frac{d}{dt}\langle P_j\rangle = & \, 2\overline{\operatorname{Re} Q_{j, j+1}} \bigg[ 2(\Delta+J)(\Delta-J)\langle P_j\rangle \\
& + (J+\Delta)^2\langle P_{j+1}\rangle + (J-\Delta)^2\langle P_{j-1}\rangle \bigg].
\end{split}
\end{equation}
Replacing the quasi-periodic potential in Eq.~(\ref{12}) with its spatial average yields:
\begin{equation}\label{ave}
\begin{aligned}
\overline{\operatorname{Re}\left(Q_{j, j+1}\right)}&\approx\frac{\sigma^{2} \theta^{2}}{\sigma^{4}+4 W^{2} \theta^{4} \sin ^{2}(\pi \beta) \overline{\sin ^{2}(\pi \beta(2 j+1))}}\\
&=\frac{\sigma^{2} \theta^{2}}{\sigma^{4}+2 W^{2} \theta^{4} \sin ^{2}(\pi \beta)}, 
\end{aligned}
\end{equation}
where we used $\overline{\sin ^{2}(\pi \beta(2 j+1))}={1}/{2}$.

Interestingly, as $t$ approaches infinity, $\overline{\operatorname{Re}(Q_{j,j+1})}$ displays a non-monotonic dependence on the noise strength $\sigma$. This behavior provides a natural explanation for the non-monotonic relaxation dynamics and the reversal of $\tau_\mathrm{relax}$ observed numerically in Fig.~\ref{fig2}. For a comprehensive discussion on the quantitative validity of these analytical approximations, including their theoretical limitations and asymptotic scaling in the extreme strong-noise limit, we refer readers to Sec. VI of the  SM.~\cite{fn}.

\subsection{Noise-induced opening of a point gap}
For the master equation in Eq.~(\ref{11}), we define the probability vector 
\begin{equation}
\langle\vec{P}\rangle \equiv \left[ \langle P_{1} \rangle, \langle P_{2} \rangle, \cdots, \langle P_{L} \rangle \right]^{T},
\end{equation}
such that  the noise master equation can be written compactly as 
\begin{equation}
    \frac{d}{dt}\langle\vec{P}\rangle = \hat{M} \langle\vec{P}\rangle. 
\end{equation}
Under periodic boundary conditions (PBCs), and using the long-time form of $\mathrm{Re} Q_{j,j+1}$ given in Eq.~(\ref{12}), the eigenvalue problem for the noisy dynamics reduces to 
\begin{equation}
    \hat{M} \langle\vec{P}\rangle = \lambda \langle\vec{P}\rangle.\label{master_eq}
\end{equation}
Fig.~\ref{fig3} compares the spectra of the master operator $\hat{M}$ with $\sigma=10$ and the static Hamiltonian ($\sigma=0$) 
under PBC for parameters $(W, J, \Delta) = (10, 0.5, 1.5)$. Notably, the static eigenenergies $E$ and the master equation eigenvalues $\lambda$ represent distinct physical quantities, governing coherent phase evolution and noise dynamics, respectively. Despite this fundamental difference, their topological structures provide a direct contrast: while the static Hamiltonian exhibits no point gap due to quasi-periodic localization, the introduction of noise results in a pronounced reopening of a point gap in the spectrum of $\hat{M}$. This noise-induced spectral winding serves as the topological origin for the resurrection of the DSE \cite{PhysRevLett.132.046602,PhysRevLett.125.126402}.
 Thus, our results provide direct theoretical evidence that OU noise restores the DSE that was initially suppressed by quasi-periodic potential-induced localization.
\begin{figure}
    \centering
    \includegraphics[width=1.1\linewidth]{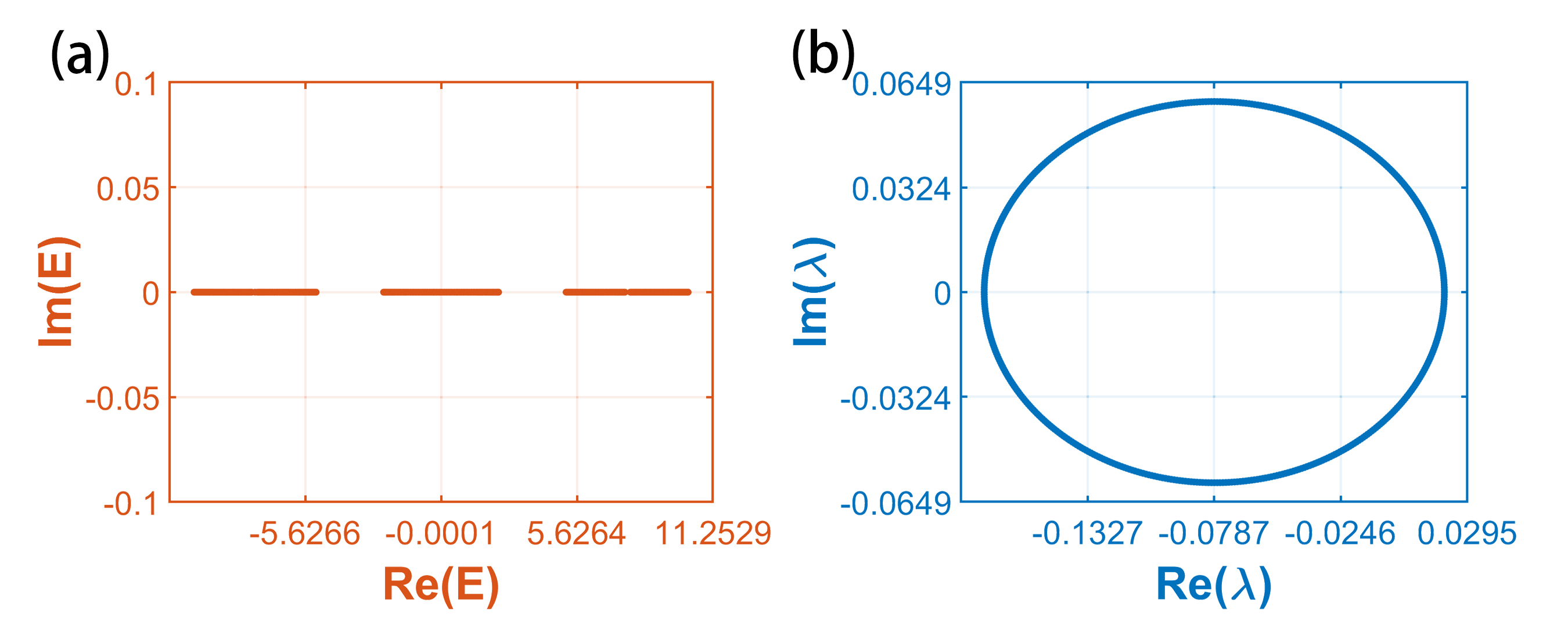}
    \caption{Complex energy spectra under PBC for the static Hamiltonian in Eq.~\eqref{Ham} (orange) (a) and the noise master equation in Eq.~\eqref{master_eq} with $\sigma=10$ (blue) (b). The remaining parameters are $(W, J, \Delta) = (10, 0.5, 1.5)$, and the system size is $L = 100$ with a PBC imposed. The introduction of noise generates a finite-area spectral loop, indicating the reopening of a point gap that dictates the emergence of DSE under OBC. }
    \label{fig3}
\end{figure}

\subsection{Probability density equation under continuous limit}
To elucidate the scaling behavior of wave-packet dynamics, we take the continuum limit of the discrete probability evolution equation in Eq.~(\ref{discrete}). Specifically, replacing probability amplitudes by their coarse-grained density representation (with $a$ being the intersite distance), 
\begin{equation}
\begin{aligned}
\left\langle P_{j}(t)\right\rangle & \Rightarrow\rho(x, t) a, \\
\left\langle P_{j \pm 1}\right\rangle & \Rightarrow\rho(x \pm a, t) a \approx a\left[\rho \pm a \frac{\partial \rho}{\partial x}+\frac{a^{2}}{2} \frac{\partial^{2} \rho}{\partial x^{2}}\right],
\end{aligned}
\end{equation}
we arrive at a drift-diffusion-reaction equation of the form \cite{10.1063/1.4936201,markowich2012semiconductor}
\[
\frac{\partial \rho(x,t)}{\partial t} = S \rho(x,t) + v \frac{\partial \rho(x,t)}{\partial x} + D \frac{\partial^2 \rho(x,t)}{\partial x^2},
\]
where the coefficients in the long-time regime are proportional to $\overline{\mathrm{Re}(Q_{j,j+1})}$ from Eq.~(\ref{ave}):
\begin{equation}\label{eq13}
\begin{aligned}
S &= 8 \overline{\mathrm{Re}(Q_{j,j+1})}\, \Delta^2, \\
v &= 8 \overline{\mathrm{Re}(Q_{j,j+1})}\, J \Delta a, \\
D &= 2 \overline{\mathrm{Re}(Q_{j,j+1})}\, (J^2 + \Delta^2) a^2.
\end{aligned}
\end{equation}
Here, $S$ is the source term, $v$ is the drift velocity, and $D$ is the diffusion coefficient.

For an initially localized wave packet, the solution exhibits:
\begin{equation}\label{eq13}
\begin{aligned}
\Delta X\left( t\right) = v t, \quad \Sigma(t) = \sqrt{2 D t},
\end{aligned}
\end{equation}
corresponding to ballistic drift and diffusive spreading at long times. 

In constrast, in the short-time regime where $\overline{\mathrm{Re}(Q_{j,j+1})} = \mathrm{Re}(Q_{j,j+1}) \sim t$, one finds:
\begin{equation}\label{eq13}
\begin{aligned}
\Delta X\left( t\right)  =4 J \Delta a t^{2},  \quad
\Sigma(t)  =\sqrt{2\left(J^{2}+\Delta^{2}\right)} a t,
\end{aligned}
\end{equation}
consistent with the initial ballistic expansion observed numerically. 
The scaling crossover between short- and long-time reveal two distinct dynamical characteristics: an initial ballistic regime dominated by the system's intrinsic non-Hermitian dynamics, and a subsequent diffusive regime emerging at long times due to the cumulative effect of noise. Therefore, our theoretical framework successfully accounts for the essential features of the noise-resurrected DSE.

\begin{table}[h]
\centering
\caption{Dynamical scaling for different time regimes.}
\label{tab:dynamical_scaling}
\begin{tabular}{c|c|c}
\hline
\hline
Time regime & $\Delta X\left( t\right)$ & $\Sigma(t)$ \\
\hline
$t \rightarrow 0$ & $4 J \Delta a t^{2}$ & $\sqrt{2}\left(J^{2}+\Delta^{2}\right)^{1/2}  a t$ \\
\hline
$t \rightarrow \infty$ & $v t$ & $\sqrt{2 D t}$ \\
\hline
\hline
\end{tabular}
\end{table}

\section{CONCLUSION}
We have demonstrated that OU-type noise can revive the DSE in non-Hermitian quasiperiodic systems, even in the deeply localized regime where the skin effect is otherwise suppressed.
Using perturbative analysis, we showed that the noise effectively maps the non-Hermitian Schr\"{o}dinger equation onto a non-reciprocal master equation, which in turn develops a noise-induced point gap in its complex spectrum. This mechanism provides a unified explanation for the restoration of DSE, the crossover between short- and long-time scaling of wave-pocket dynamics, and the non-monotonic dependence of the relaxation behavior on noise strength. 

Our findings establish noise as a robust tuning parameter for non-Hermitian transport in quasiperiodic lattices.  Furthermore, our derivation is general and not limited to quasiperiodic systems, which implies that the noise recovery DSE is also manifest in disordered non-Hermitian Anderson systems. Future directions include extending this versatile framework to recent theoretical developments in non-Hermitian physics \cite{1111,PhysRevB.111.094109,kcm2-2mz4,5ksl-tjjm,PhysRevLett.134.196302}, as well as experimental implementations in photonic, cold-atom, electrical-circuit, and acoustic platforms.

\begin{acknowledgments}
This work is supported by the National Key R\&D Program of China (Grant No. 2021YFA1401600), the National Natural Science Foundation of China (Grant No. 12474056). The work was carried out at the National Supercomputer Center in Tianjin, and the calculations were performed on Tianhe new generation supercomputer. The high-performance computing platform of Peking University supported the computational resources.
\end{acknowledgments}

\bibliography{reference}

\end{document}


\title{Supplementary Material for "Noise-Induced Resurrection of Dynamical Skin Effects in Quasiperiodic Non-Hermitian Systems"}

\author{Wuping Yang}
\affiliation{School of Physics, Peking University, Beijing 100871, China}

\author{H. Huang}%
 \email[Corresponding author: ]{huanghq07@pku.edu.cn}
\affiliation{School of Physics, Peking University, Beijing 100871, China}
\affiliation{Collaborative Innovation Center of Quantum Matter, Beijing 100084, China}
\affiliation{Center for High Energy Physics, Peking University, Beijing 100871, China}

\date{\today}

\maketitle
\tableofcontents
\clearpage

\section{An introductory introduction to Ornstein-Uhlenbeck process}

\subsection{Wiener Process}
To fully introduce the OU process, it is necessary to first introduce the Wiener process, which is the most basic stochastic process. 

\textbf{Definition:} A stochastic process $\{W(t), t \geqslant 0\}$ is called a Wiener process if it satisfies:
\begin{enumerate}[label=(\arabic*)]
    \item $W(0)=0$;
    \item It has stationary and independent increments;
    \item For $t>0$, $W(t) \sim \mathcal{N}(0, t)$.
\end{enumerate}

Studying the increment $dW_t \equiv W(t + dt) - W(t) \sim \mathcal{N}(0, dt)$ of the Wiener process is essential for our subsequent discussion. A notable property is that $dW_t^2 = dt$. Below we will prove this property.

\subsubsection*{Proof of $dW_t^2 = dt$}

First, we compute the expected value of $dW_t^2$:
\begin{equation}
    \mathbb{E}[dW_t^2] = \operatorname{Var}[dW_t] + (\mathbb{E}[dW_t])^2 = dt + 0 = dt.
\end{equation}

Next, we calculate the variance of $dW_t^2$:
\begin{equation}
    \operatorname{Var}[dW_t^2] = \mathbb{E}[dW_t^4] - (\mathbb{E}[dW_t^2])^2.
\end{equation}

Since $dW_t \sim \mathcal{N}(0, dt)$, we use the Gaussian moment property:
\begin{equation}
    \begin{split}
        \mathbb{E}[dW_t^2] &= dt ,\\
        \mathbb{E}[dW_t^4] &= 3(dt)^2.
    \end{split}
\end{equation}
Substituting (S3) into (S2):
\begin{equation}
    \operatorname{Var}[dW_t^2] = 3(dt)^2 - (dt)^2 = 2(dt)^2 = 0 .
\end{equation}
Therefore, it has been established that $dW_t^2 = dt$ holds to first order in $dt$.

\subsubsection*{Itô's Lemma}

The relation $dW_t^2 = dt$ leads to non-trivial effects in stochastic calculus. When differentiating a stochastic process $X_t$, we must account for terms up to $\mathcal{O}(dW_t^2)$ to capture all contributions of order $\mathcal{O}(dt)$. This is the fundamental insight behind \textbf{Itô's Lemma}. And if we want to explain Ito's lemma clearly, we need to introduce the concept of Itô process.

\textbf{Definition:}
A stochastic process $X_t$ is called an \textit{Itô process} if it satisfies:
\begin{equation}
dX_t = a_t dt + b_t dW_t,
\label{eq:ito_process}
\end{equation}
where $a_{t}$ and $b_{t}$ are real variables related to time $t$, respectively. And $W_t$ is a standard Wiener process.
    
For any twice continuously differentiable function $Y_t = f(t,X_t)$ where $X_t$ is an Itô process, we obtain its stochastic differential through second-order Taylor expansion:

\begin{equation}
\begin{split}
dY_t &= \frac{\partial f}{\partial t}dt + \frac{\partial f}{\partial X_t}dX_t \\
&\quad + \frac{1}{2}\left(\frac{\partial^2 f}{\partial t^2}(dt)^2 + 2\frac{\partial^2 f}{\partial t \partial X_t}dt\,dX_t + \frac{\partial^2 f}{\partial X_t^2}(dX_t)^2\right).
\label{eq:taylor_expansion}
\end{split}
\end{equation}

Substituting the Itô process $dX_t = a_t dt + b_t dW_t$ into Eq. (S6) and applying the standard Itô rules—$(dt)^2 = 0$, $dt dW_t = 0$, and $(dW_t)^2 = dt$—we collect the terms to first order in $dt$ to arrive at the final result:
\begin{equation}
dY_t = \left(\frac{\partial f}{\partial t} + a_t\frac{\partial f}{\partial X_t} + \frac{1}{2}b_t^2\frac{\partial^2 f}{\partial X_t^2}\right)dt + b_t\frac{\partial f}{\partial X_t}dW_t.
\label{eq:ito_lemma}
\end{equation}

This fundamental result constitutes Itô's Lemma, characterized by a essential feature: It contains the second-order correction term $\frac{1}{2}b_t^2\frac{\partial^2 f}{\partial X_t^2}dt$ that arises from the non-trivial quadratic variation of Wiener process. 

\subsection{Ornstein-Uhlenbeck Process}
The OU process is given by a differential equation containing $dW_t$: 
\begin{equation}
d X_{t}=\theta\left(\mu-X_{t}\right) d t+\sigma d W_{t} ,
\end{equation}
where $\theta$, $\mu$ and $\sigma$ are all real variables, and $\theta$, $\sigma$ are greater than 0. 

The parameters have distinct physical interpretations: $\theta$ is mean reversion rate, which means the speed of returning to equilibrium.
$\sigma$ is volatility coefficient, which means magnitude of random fluctuations.
$\mu$ is the long-term average value of $X_{t}$. Next, we will solve the differential equation and obtain the general solution form of $X_{t}$. 

Firstly, we transform the differential equation into the following form:
\begin{equation}\label{non-homogeneous}
d X_{t}+\theta X_{t} d t=\theta \mu d t+\sigma d W_{t}.
\end{equation}

Examine the homogeneous equation case:
\begin{equation}
dX_{t} + \theta X_{t} dt = 0.
\end{equation}

This ordinary differential equation admits the following general solution:
\begin{align}
\frac{dX_{t}}{X_{t}} &= -\theta dt.
\end{align}
By integrating both sides of the equation simultaneously and simplifying, we obtain:
\begin{align}
X_{t} &= Ce^{-\theta t} \label{eq:sol},
\end{align}
where $C$ is the constant of integration determined by initial conditions.

To solve the non-homogeneous stochastic differential equation \eqref{non-homogeneous}, we employ the method of variation of parameters. Assume the general solution takes the form:
\begin{equation}
X_{t} = C_{t} e^{-\theta t},
\label{eq:ansatz}
\end{equation}
where the constant $C$ from the homogeneous solution is now treated as a time-dependent function $C_t$.

Applying Itô's lemma to differentiate the ansatz \eqref{eq:ansatz}, where $f(t,C_t) = C_t e^{-\theta t}$, we obtain:
\begin{equation}
dX_t = \frac{\partial f}{\partial t} dt + \frac{\partial f}{\partial C_t} dC_t + \frac{1}{2} \frac{\partial^2 f}{\partial C_t^2} (dC_t)^2.
\label{eq:ito_application}
\end{equation}

Computing the partial derivatives explicitly, we find that the second-order term vanishes due to the linearity of $f$ in $C_t$. Substituting results into \eqref{eq:ito_application} yields the simplified differential form:
\begin{equation}
dX_t = e^{-\theta t} dC_t - \theta C_t e^{-\theta t} dt.
\label{eq:diff_ansatz}
\end{equation}

Substituting \eqref{eq:diff_ansatz} into the original equation (18) and simplifying gives:
\begin{equation}
dC_{t} = \theta \mu e^{\theta t} dt + \sigma e^{\theta t} dW_{t}.
\label{eq:dCt}
\end{equation}

Integrating both sides of \eqref{eq:dCt} from 0 to $t$ produces:
\begin{equation}
C_{t} = C_{0} + \mu(e^{\theta t} - 1) + \sigma \int_{0}^{t} e^{\theta s} dW_{s}
\label{eq:Ct_solution}
\end{equation}

Finally, substituting \eqref{eq:Ct_solution} back into our ansatz \eqref{eq:ansatz} yields the complete solution:
\begin{equation}
X_{t} = X_{0} e^{-\theta t} + \mu(1 - e^{-\theta t}) + \sigma \int_{0}^{t} e^{-\theta(t-s)} dW_{s} ,
\label{eq:final_solution}
\end{equation}
where the first two items are the deterministic part. The third term is random integration, which follows a Gaussian distribution. Therefore, $X_{t}$ is also a random variable that follows a Gaussian distribution.

After obtaining the analytical solution for $X_{t}$, computing the expectation and covariance of the process becomes straightforward. 

\subsubsection*{Expectation Calculation}
Taking the expectation of both sides of the solution:
\begin{equation}
\mathbb{E}[X_{t}] = \mathbb{E}\left[X_{0} e^{-\theta t} + \mu\left(1-e^{-\theta t}\right) + \sigma \int_{0}^{t} e^{-\theta(t-s)} dW_{s}\right].
\end{equation}
Since $\mathbb{E}[dW_s] = 0$, the stochastic integral vanishes, yielding:
\begin{equation}
\mathbb{E}[X_{t}] = X_{0} e^{-\theta t} + \mu\left(1-e^{-\theta t}\right).
\label{eq:expectation}
\end{equation}
Eq. \eqref{eq:expectation} shows that as $t \to \infty$, $\mathbb{E}[X_{t}] \to \mu$, which identifies $\mu$ as the mean-reversion level.

\subsubsection*{Covariance Calculation}
The covariance is defined as:
\begin{equation}
\operatorname{Cov}[X_{s}, X_{t}] \equiv \mathbb{E}\left[\left(X_{s}-\mathbb{E}[X_{s}]\right)\left(X_{t}-\mathbb{E}[X_{t}]\right)\right].
\end{equation}

Using $X_{t} - \mathbb{E}[X_{t}] = \sigma \int_{0}^{t} e^{-\theta(t-u)} dW_{u}$, we obtain:
\begin{equation}
\operatorname{Cov}[X_{s}, X_{t}] = \sigma^{2} \mathbb{E}\left[\left(\int_{0}^{s} e^{-\theta(s-u)} dW_{u}\right)\left(\int_{0}^{t} e^{-\theta(t-v)} dW_{v}\right)\right].
\end{equation}

Applying the property $\mathbb{E}[dW_u dW_v] = \delta_{uv} du$, we can have:
\begin{align}
\operatorname{Cov}[X_{s}, X_{t}] &= \sigma^{2} \int_{0}^{\min(s, t)} e^{-\theta(s-u)} e^{-\theta(t-u)} du \nonumber \\
&= \frac{\sigma^{2}}{2\theta} e^{-\theta(s+t)} \left(e^{2\theta \min(s, t)} - 1\right) \nonumber \\
&= \frac{\sigma^{2}}{2\theta} \left(e^{-\theta|s-t|} - e^{-\theta(s+t)}\right).
\label{eq:covariance}
\end{align}

The asymptotic covariance for large $s$ and $t$ becomes:
\begin{equation}
\operatorname{Cov}[X_{s}, X_{t}] \sim \frac{\sigma^{2}}{2\theta} e^{-\theta|s-t|}.
\label{eq:asymptotic_cov}
\end{equation}

\subsection{Proof of a useful identity for gaussian random variables}

From \eqref{eq:final_solution}, it can be seen that the random numbers generated by the OU process follow a Gaussian distribution. We now introduce and prove an important identity for Gaussian random variables, which will be utilized in later derivations.

Suppose  $X$ is a Gaussian random variable ($X \sim \mathcal{N}(\mu, \sigma^2)$). Then the following identity holds:
\begin{equation}
\mathbb{E}\left[e^{i X}\right] = e^{i \mathbb{E}[X] - \frac{1}{2} \operatorname{Var}(X)} = e^{i \mu - \frac{\sigma^{2}}{2}}.
\end{equation}

The proof proceeds as follows. First, we express the expectation directly as an integral using the probability density function of the Gaussian distribution:
\begin{equation}
\mathbb{E}\left[e^{i X}\right] = \int_{-\infty}^{\infty} e^{i x} \frac{1}{\sqrt{2 \pi \sigma^{2}}} e^{-\frac{(x-\mu)^{2}}{2 \sigma^{2}}} dx.
\end{equation}
To simplify the exponent in the integrand, we perform the following algebraic manipulation:
\begin{equation}
\begin{aligned}
i x - \frac{(x-\mu)^{2}}{2 \sigma^{2}} 
&= -\frac{1}{2 \sigma^{2}}\left(x^{2} - 2(\mu + i \sigma^{2}) x + \mu^{2}\right) \\
&= -\frac{\left(x - (\mu + i \sigma^{2})\right)^{2}}{2 \sigma^{2}} + i \mu - \frac{\sigma^{2}}{2}.
\end{aligned}
\end{equation}
This simplification allows us to rewrite the expectation as:
\begin{equation}
\mathbb{E}\left[e^{i X}\right] = e^{i \mu - \frac{\sigma^{2}}{2}} \cdot \frac{1}{\sqrt{2 \pi \sigma^{2}}} \int_{-\infty}^{\infty} e^{-\frac{\left(x - (\mu + i \sigma^{2})\right)^{2}}{2 \sigma^{2}}} dx.
\end{equation}
Using the standard Gaussian integral formula
$\int_{-\infty}^{\infty} e^{-\frac{\left(x - (\mu + i \sigma^{2})\right)^{2}}{2 \sigma^{2}}} dx = \sqrt{2 \pi \sigma^{2}}$
we obtain the final result:
\begin{equation}
\mathbb{E}\left[e^{i X}\right] = e^{i \mu - \frac{\sigma^{2}}{2}}.
\end{equation}

Recall that for Gaussian distributions, $\mathbb{E}[X] = \mu$ and $\operatorname{Var}(X) = \sigma^{2} = \mathbb{E}[X^{2}] - \mu^{2}$. Therefore, we can alternatively express the result as:
\begin{equation}
\mathbb{E}\left[e^{i X}\right] = e^{i \mu - \frac{\sigma^{2}}{2}} = e^{i \mathbb{E}[X] - \frac{1}{2} \operatorname{Var}(X)}.
\end{equation}
If $\mu=0$, then the above equation can be further simplified as:
\begin{equation}\label{S31}
\begin{array}{l}
\mathbb{E}\left[e^{i X}\right]=e^{-\frac{1}{2} \sigma^{2}}=e^{-\frac{1}{2}\mathbb{E}\left[X^{2}\right]} .
\end{array}
\end{equation}

\section{Establishing the master equation using perturbation theory}

\subsection{The master equation under the clean limit}
Firstly, we analyze the Hatano-Nelson model \cite{PhysRevB.56.8651,PhysRevLett.77.570} with OU noise through perturbation theory \cite{PhysRevE.79.050105}. The system dynamics are governed by:
\begin{equation}\label{S32}
i \partial_t A_j(t) = (J+\Delta)A_{j+1}(t) + (J-\Delta)A_{j-1}(t) + \xi(j,t)A_j(t),
\end{equation} 
where $A_j(t)$ is the wavefunction amplitude at site $j$, $\xi(j,t)$ is OU noise with $\langle \xi(j,t)\xi(j',t+\tau) \rangle =\frac{\sigma^{2}}{2\theta} e^{-\theta|\tau|}\delta_{jj'}$, $J$ is the hopping amplitude, and $\Delta$ quantifies non-reciprocity.

In the strong noise regime ($\sigma \gg J, \Delta$), we perform a perturbative expansion of $A_j(t)$ in the form $A_j(t) = A_j^{(0)}(t) + A_j^{(1)}(t)$. The zeroth-order solution satisfies:
\begin{equation}
i \partial_t A_j^{(0)}(t) = \xi(j,t) A_j^{(0)}(t),
\end{equation}
while the first-order correction obeys:
\begin{equation}
\begin{aligned}
i \partial_t A_j^{(1)}(t) &= J\left(A_{j+1}^{(0)}(t) + A_{j-1}^{(0)}(t)\right) \\
&\quad + \Delta\left(A_{j+1}^{(0)}(t) - A_{j-1}^{(0)}(t)\right) + \xi(j,t) A_j^{(1)}(t).
\end{aligned}\label{differential}
\end{equation}

Defining $\phi_{j}(t) \equiv \int_{0}^{t} \xi(j,\tau) d\tau$, we solve the differential equatio \eqref{differential} to obtain:
\begin{equation}
\begin{aligned}
\frac{d}{dt}\left(A_{j}^{(1)} e^{i \phi_{j}(t)}\right) &= e^{i \phi_{j}(t)}\left(\frac{J}{i}\right)\left(A_{j+1}^{(0)} + A_{j-1}^{(0)}\right) \\
&\quad + e^{i \phi_{j}(t)}\left(\frac{\Delta}{i}\right)\left(A_{j+1}^{(0)} - A_{j-1}^{(0)}\right).
\end{aligned}
\end{equation}

Integrating both sides yields:
\begin{equation}
\begin{aligned}\label{S36}
A_{j}^{(1)} &= e^{-i \phi_{j}(t)} \int_{0}^{t}\left(\frac{J}{i}\right) e^{i \phi_{j}(\tau)}\left[A_{j+1}^{(0)}(\tau) + A_{j-1}^{(0)}(\tau)\right] d\tau \\
&\quad + e^{-i \phi_{j}(t)} \int_{0}^{t}\left(\frac{\Delta}{i}\right) e^{i \phi_{j}(\tau)}\left[A_{j+1}^{(0)}(\tau) - A_{j-1}^{(0)}(\tau)\right] d\tau.
\end{aligned}
\end{equation}

Considering the time derivative of the probability density $P_j \equiv |A_j|^2 = A_j^* A_j$:
\begin{equation}
\frac{d}{dt}P_j = \frac{d}{dt}A_j^* \cdot A_j + A_j^* \cdot \frac{d}{dt}A_j,
\end{equation}
and using utilizing (\ref{S32}) and its conjugated forms, we obtain:
\begin{equation} \label{S38}
\begin{aligned}
\frac{d}{dt}P_j &= iJ\left(A_{j+1}^* A_j + A_{j-1}^* A_j - A_j^* A_{j+1} - A_j^* A_{j-1}\right) \\
&\quad + i\Delta\left(A_{j+1}^* A_j - A_{j-1}^* A_j - A_j^* A_{j+1} + A_j^* A_{j-1}\right).
\end{aligned}
\end{equation}

Using the identity $z - z^* = 2i\operatorname{Im}(z)$ for $z \in \mathbb{C}$, (\ref{S38}) simplifies to:
\begin{equation} \label{S39}
\begin{aligned}
\frac{d}{dt}P_j &= 2J\operatorname{Im}\left(A_j^* A_{j+1} + A_j^* A_{j-1}\right) \\
&\quad + 2\Delta\operatorname{Im}\left(A_j^* A_{j+1} - A_j^* A_{j-1}\right).
\end{aligned}
\end{equation}

Taking the noise average, retaining only first-order terms and noting that noise at different sites is uncorrelated ($\langle A_i^{(0)*}(t)A_j^{(0)}(t)\rangle = 0$ for $i \neq j$), we have:
\begin{equation}\label{S40}
\begin{aligned}
\left\langle\frac{d}{dt}P_j\right\rangle &\approx 2J\left\langle\operatorname{Im}\left[A_j^{(0)*}A_{j+1}^{(1)} + A_j^{(1)*}A_{j+1}^{(0)} + A_j^{(0)*}A_{j-1}^{(1)} + A_j^{(1)*}A_{j-1}^{(0)}\right]\right\rangle \\
&\quad + 2\Delta\left\langle\operatorname{Im}\left[A_j^{(0)*}A_{j+1}^{(1)} + A_j^{(1)*}A_{j+1}^{(0)} - A_j^{(0)*}A_{j-1}^{(1)} - A_j^{(1)*}A_{j-1}^{(0)}\right]\right\rangle.
\end{aligned}
\end{equation}

Substituting the expression for $A_j^{(1)}$, we can evaluate terms like $\langle A_j^{(0)*}A_{j+1}^{(1)} \rangle$:
\begin{equation}
\begin{aligned}
\langle A_j^{(0)*}A_{j+1}^{(1)} \rangle &= \langle A_j^{(0)*}(t)e^{-i\phi_{j+1}(t)} \int_0^t \left(\frac{J}{i}\right)e^{i\phi_{j+1}(\tau)}A_j^{(0)}(\tau)d\tau  \rangle \\
&\quad - \langle A_j^{(0)*}(t)e^{-i\phi_{j+1}(t)} \int_0^t \left(\frac{\Delta}{i}\right)e^{i\phi_{j+1}(\tau)}A_j^{(0)}(\tau)d\tau  \rangle.
\end{aligned}
\end{equation}

Using $A_j^{(0)}(t) = A_j^{(0)}(0)e^{-i\phi_j(t)}$, $A_j^{(0)*}(t) = A_j^{(0)}(0)e^{i\phi_j(t)}$ and $\left\langle P_{j}\right\rangle \approx\left|A_{j}^{(0)}\right|^{2}$ ,we find:
\begin{equation}
\begin{aligned}
\langle A_j^{(0)*}A_{j+1}^{(1)}  \rangle &\approx \left(\frac{J}{i}\right)\langle P_j\rangle \langle \int_0^t e^{i[\phi_j(t)-\phi_j(\tau)]}e^{-i[\phi_{j+1}(t)-\phi_{j+1}(\tau)]}d\tau\rangle \\
&\quad - \left(\frac{\Delta}{i}\right)\langle P_j\rangle \langle\int_0^t e^{i[\phi_j(t)-\phi_j(\tau)]}e^{-i[\phi_{j+1}(t)-\phi_{j+1}(\tau)]}d\tau\rangle.
\end{aligned}
\end{equation}

Defining the correlation integrals:
\begin{equation}
\begin{aligned}
Q(t) &\equiv \left\langle\int_0^t e^{i[\phi_j(t)-\phi_j(\tau)]}e^{-i[\phi_{j+1}(t)-\phi_{j+1}(\tau)]}d\tau\right\rangle, \\
Q^*(t) &\equiv \left\langle\int_0^t e^{i[\phi_j(t)-\phi_j(\tau)]}e^{-i[\phi_{j-1}(t)-\phi_{j-1}(\tau)]}d\tau\right\rangle,
\end{aligned}
\end{equation}
we obtain:
\begin{equation}
\langle A_j^{(0)*}A_{j+1}^{(1)}\rangle = (-iJ)\langle P_j\rangle Q(t) + (i\Delta)\langle P_j\rangle Q(t).
\end{equation}

Similarly, the other terms are evaluated to:
\begin{equation}
\begin{aligned}
\langle A_j^{(1)*}A_{j+1}^{(0)}\rangle &= (iJ)\langle P_{j+1}\rangle Q(t) + (i\Delta)\langle P_{j+1}\rangle Q(t), \\
\langle A_j^{(0)*}A_{j-1}^{(1)}\rangle &= (-iJ)\langle P_j \rangle Q^*(t) + (-i\Delta)\langle P_j \rangle Q^*(t), \\
\langle A_j^{(1)*}A_{j-1}^{(0)}\rangle &= (iJ)\langle P_{j-1}\rangle Q^*(t) - (i\Delta)\langle P_{j-1}\rangle Q^*(t).
\end{aligned}
\end{equation}

Substituting the above result into Eq. \eqref{S40} yields the effective noise master equation:
\begin{equation}\label{S46}
\frac{d}{dt}\langle P_j\rangle = 2\operatorname{Re}(Q)\left[2(\Delta+J)(\Delta-J)\langle P_j\rangle + (J+\Delta)^2\langle P_{j+1}\rangle + (J-\Delta)^2\langle P_{j-1}\rangle\right].
\end{equation}

When the non-reciprocal term $\Delta = 0$, \eqref{S46} reduces to the diffusion equation. Larger $\operatorname{Re}(Q)$ indicate better transport properties. For $\Delta \neq 0$, we expect $\operatorname{Re}(Q)$ still serves as a good measure of system transport (In fact, we will also see in the subsequent derivation that $\operatorname{Re}(Q)$ will be directly correlated with observable quantities in transportation). 
Notably, in the final expression \eqref{S46}, only the prefactor $\operatorname{Re}(Q)$ depends on noise, indicating that noise does not affect the emergence of DSE.

\subsection{Master equation under 
onsite potential that breaks translational symmetry}
When introducing an onsite potential $\varepsilon _{j}$ that can break the translational symmetry to system, the Schr\"{o}dinger equation modifies to:
\begin{equation}
i \partial_{t} A_{j}(t) = (J+\Delta) A_{j+1}(t) + (J-\Delta) A_{j-1}(t) + \xi(j,t) A_{j}(t) + \varepsilon _{j} A_{j}(t).
\label{eq:quasiperiodic}
\end{equation}
Following the previous perturbation approach, the zeroth-order solution now becomes:
\begin{equation}
A_{j}^{(0)}(t) = A_{j}^{(0)}(0) e^{-i \varepsilon _{j} t - i \int_{0}^{t} \xi(j,\tau) d\tau}.
\label{eq:zeroth_order_quasi}
\end{equation}
The first-order correction maintains the same form as Eq.~(\ref{S36}), but with a modified phase factor:
\begin{equation}
\phi_{j}(t) = \varepsilon _{j} t + \int_{0}^{t} \xi(j,\tau) d\tau.
\label{eq:phase_def}
\end{equation}
This requires redefining the $Q$-functions with explicit site-dependence:
\begin{align}
Q_{j, j+1} &\equiv \langle\int_{0}^{t} e^{i[\phi_{j}(t)-\phi_{j}(\tau)]} e^{-i[\phi_{j+1}(t)-\phi_{j+1}(\tau)]} d\tau \rangle \label{eq:Q_plus}, \\
Q_{j, j-1} &\equiv \langle\int_{0}^{t} e^{i[\phi_{j}(t)-\phi_{j}(\tau)]} e^{-i[\phi_{j-1}(t)-\phi_{j-1}(\tau)]} d\tau \rangle .\label{eq:Q_minus}
\end{align}
The correlation terms consequently become:
\begin{equation}\label{S52}
\begin{array}{l}
\langle A_{j}^{(0) *} A_{j+1}^{(1)}\rangle=(-i J) \langle P_{j}\rangle   Q_{j, j+1}(t)+(i \Delta) \langle P_{j}\rangle   Q_{j, j+1}(t), \\
\langle A_{j}^{(1) *} A_{j+1}^{(0)}\rangle=(i J) \langle P_{j+1}\rangle  Q_{j, j+1}(t)+(i \Delta) \langle P_{j+1}\rangle   Q_{j, j+1}(t), \\
\langle A_{j}^{(0) *} A_{j-1}^{(1)}\rangle=(-i J) \langle P_{j}\rangle   Q_{j, j-1}(t)+(-i \Delta) \langle P_{j}\rangle   Q_{j, j-1}(t), \\
\langle A_{j}^{(1) *} A_{j-1}^{(0)}\rangle=(i J) \langle P_{j-1}\rangle   Q_{j, j-1}(t)+(-i \Delta) \langle P_{j-1}\rangle   Q_{j, j-1}(t).
\end{array}
\end{equation}
Remarkably, the onsite potential preserves the form of the probability equation (\ref{S38}). Substituting Eq.~(\ref{S52}) into Eq.~(\ref{S38}), after straightforward but lengthy algebra, we can have the noise master equation:
\begin{equation}\label{S53}
\begin{aligned}
\frac{d}{d t}\left\langle P_{j}\right\rangle= & 2\left\{\operatorname{Re}\left( Q_{j, j+1}+Q_{j, j-1}\right)(\Delta+J)(\Delta-J)\left\langle P_{j}\right\rangle+\right. \\
& \left.\operatorname{Re}\left( Q_{j, j+1}\right)(\Delta+J)^{2}\left\langle P_{j+1}\right\rangle+\operatorname{Re}\left( Q_{j, j-1}\right)(\Delta-J)^{2}\left\langle P_{j-1}\right\rangle\right\}.
\end{aligned}
\end{equation}

\subsection{$\operatorname{Re} Q_{j, j+1}$ in quasi-periodic potentials}  
We now consider a translational symmetry-breaking potential characterized by a quasi-periodic potential $\varepsilon_j = W\cos(2\pi\beta j)$. $\operatorname{Re} Q_{j, j+1}$ can be expanded as:
\begin{equation}\label{S54}
\begin{aligned}
\operatorname{Re} Q_{j, j+1} 
&= \left\langle \operatorname{Re}\left[\int_{0}^{t} e^{i(\varepsilon_j - \varepsilon_{j+1})(t-\tau)} e^{i \int_{\tau}^{t} \xi(j,s) ds} e^{-i \int_{\tau}^{t} \xi(j+1,s) ds} d\tau\right]\right\rangle \\
&= \int_{0}^{t} \cos\left[(\varepsilon_j - \varepsilon_{j+1})(t-\tau)\right] \left|C_\phi(t-\tau)\right|^2 d\tau,
\end{aligned}
\end{equation}
where $C_\phi(t-\tau) \equiv \langle e^{i \int_{\tau}^{t} \xi(j,s) ds}\rangle$. Crucially, since the noise at different lattice sites is independent, $C_\phi(t)$ is position-independent; thus, we drop the spatial index in $\xi(j,s)$ in the subsequent derivation.
Note that $\xi(s)$ is a random number that follows a Gaussian distribution, therefore $\int_{\tau}^{t} \xi(s) d s$ should also be a random number that follows a Gaussian distribution \footnote{The sum of two independent Gaussian random variables also follows a Gaussian distribution.}. Therefore, using Eq.~(\ref{S31}), we obtain:
\begin{equation}\label{S55}
C_\phi(t-\tau) = \left\langle e^{-i \int_{\tau}^{t} \xi(t) dt}\right\rangle = e^{-\frac{1}{2}\left\langle\left(\int_{\tau}^{t} \xi(t) dt\right)^2\right\rangle}.
\end{equation}
The argument of the exponential evaluates to:
\begin{equation}\label{S56}
\begin{aligned}
\left\langle\left(\int_{\tau}^{t} \xi(s) ds\right)^2\right\rangle 
&= \int_{\tau}^{t} \int_{\tau}^{t} \left\langle\xi(s_1)\xi(s_2)\right\rangle ds_1 ds_2 \\
& =\frac{\sigma^{2}}{2\theta}[\int_{\tau}^{t}\int_{\tau}^{s_{1}}e^{-\theta(s_{1}-s_{2})}ds_{2}ds_{1}+\int_{\tau}^{t}\int_{s_{2}}^{t}e^{-\theta(s_{2}-s_{1})}ds_{1}ds_{2}] \\
 & =\frac{\sigma^{2}}{\theta}\int_{\tau}^{t}\int_{\tau}^{s_{1}}e^{-\theta(s_{1}-s_{2})}ds_{2}ds_{1} \\
 & =\frac{\sigma^{2}}{\theta}\int_{\tau}^{t}\frac{1}{\theta}\left(1-e^{-\theta\left(s_{1}-\tau\right)}\right)ds_{1} \\
 & =\frac{\sigma^{2}}{\theta^{2}}[\Delta t-\frac{1}{\theta}(1-e^{-\theta\Delta t})]\quad (\Delta t \equiv t-\tau).
\end{aligned}
\end{equation}

Substituting Eq.~(\ref{S56}) into Eq.~(\ref{S54}) with the quasiperiodic potential $\varepsilon_j = W\cos(2\pi\beta j)$ yields:
\begin{equation}
\begin{aligned}
\operatorname{Re} Q_{j, j+1} 
&= \int_{0}^{t} \cos \big[W\big(\cos (2\pi\beta j) - \cos (2\pi\beta(j+1))\big)(t-\tau)\big] \left|C_\phi(t-\tau)\right|^2 d\tau.
\end{aligned}
\end{equation}
Examining the long-time limit behavior of $\operatorname{Re} Q_{j, j+1}$, we have:
\begin{equation}
\begin{aligned}
\operatorname{Re} Q_{j, j+1}
&=\lim _{t\rightarrow \infty }\int_{0}^{t} \cos [W(\cos (2 \pi \beta j)-\cos (2 \pi \beta(j+1))) (t-\tau)] e^{-\frac{\sigma^{2}}{\theta^{2}}\left[(t-\tau)-\frac{1}{\theta}\left(1-e^{-\theta (t-\tau)}\right)\right]} d\tau \\
&\approx \int_{0}^{\infty} \cos \big[W\big(\cos (2\pi\beta j) - \cos (2\pi\beta(j+1))\big)\Delta t\big] e^{-\frac{\sigma^{2}}{\theta^{2}}\Delta t} d(\Delta t).
\label{long_time}
\end{aligned}
\end{equation}
Employing the standard integral identity $\int_{0}^{\infty} \cos (kx) e^{-a x} \, dx = \frac{a}{a^{2}+k^{2}}$ for $a>0$, we obtain:
\begin{equation}\label{S59}
\operatorname{Re} Q_{j, j+1} = \frac{\sigma^2\theta^2}{\sigma^4+4W^2\theta^4\sin^2(\pi\beta(2j+1))\sin^2(\pi\beta)}.
\end{equation}
Here we employed the trigonometric identity $\cos x - \cos y = -2 \sin\left(\frac{x+y}{2}\right) \sin\left(\frac{x-y}{2}\right)$. 

To rewrite the expression (\ref{S53}) in (\ref{S46}) form,  we need to calculate the spatial average $\overline{\operatorname{Re} Q_{j, j+1}}$.  
Replace the part about quasicrystal potential in (\ref{S59}) directly with its spatial average value, we obtain:
\begin{equation}\label{S60}
\begin{aligned}
\overline{\operatorname{Re}\left(Q_{j, j+1}\right)}&\approx\frac{\sigma^{2} \theta^{2}}{\sigma^{4}+4 W^{2} \theta^{4} \sin ^{2}(\pi \beta) \overline{\sin ^{2}(\pi \beta(2 j+1))}}.\\
\end{aligned}
\end{equation}
By taking the continuum limit, the spatial average $\overline{\sin^{2}(\pi \beta(2j+1))}$ can be evaluated via integration as:
\begin{equation}
\begin{aligned}
\overline{\sin ^{2}(\pi \beta(2 j+1))} & =\lim _{L \rightarrow \infty}\frac{1}{L} \sum_{j=1}^{L} \sin ^{2}(\pi \beta(2 j+1)) \\
& =\lim _{L \rightarrow \infty}\frac{1}{L} \int_{0}^{L} \sin ^{2}(\pi \beta(2 x+1)) d x \\
& =\lim _{L \rightarrow \infty}\frac{1}{L} \int_{0}^{L} \frac{1-\cos (4 \pi \beta x+2 \pi \beta)}{2} d x \\
& =\lim _{L \rightarrow \infty}\frac{1}{2 L} \int_{0}^{L} d x-\frac{1}{2 L} \int_{0}^{L} \cos (4 \pi \beta x+2 \pi \beta) d x \\
& =\frac{1}{2}-\lim _{L \rightarrow \infty}\frac{1}{2 L} \cdot \frac{1}{4 \pi \beta}[\sin (4 \pi \beta L+2 \pi \beta)-\sin (2 \pi \beta)] \\
& =\frac{1}{2}.
\end{aligned}
\end{equation}

\begin{figure}
    \centering
    \includegraphics[width=0.8\linewidth]{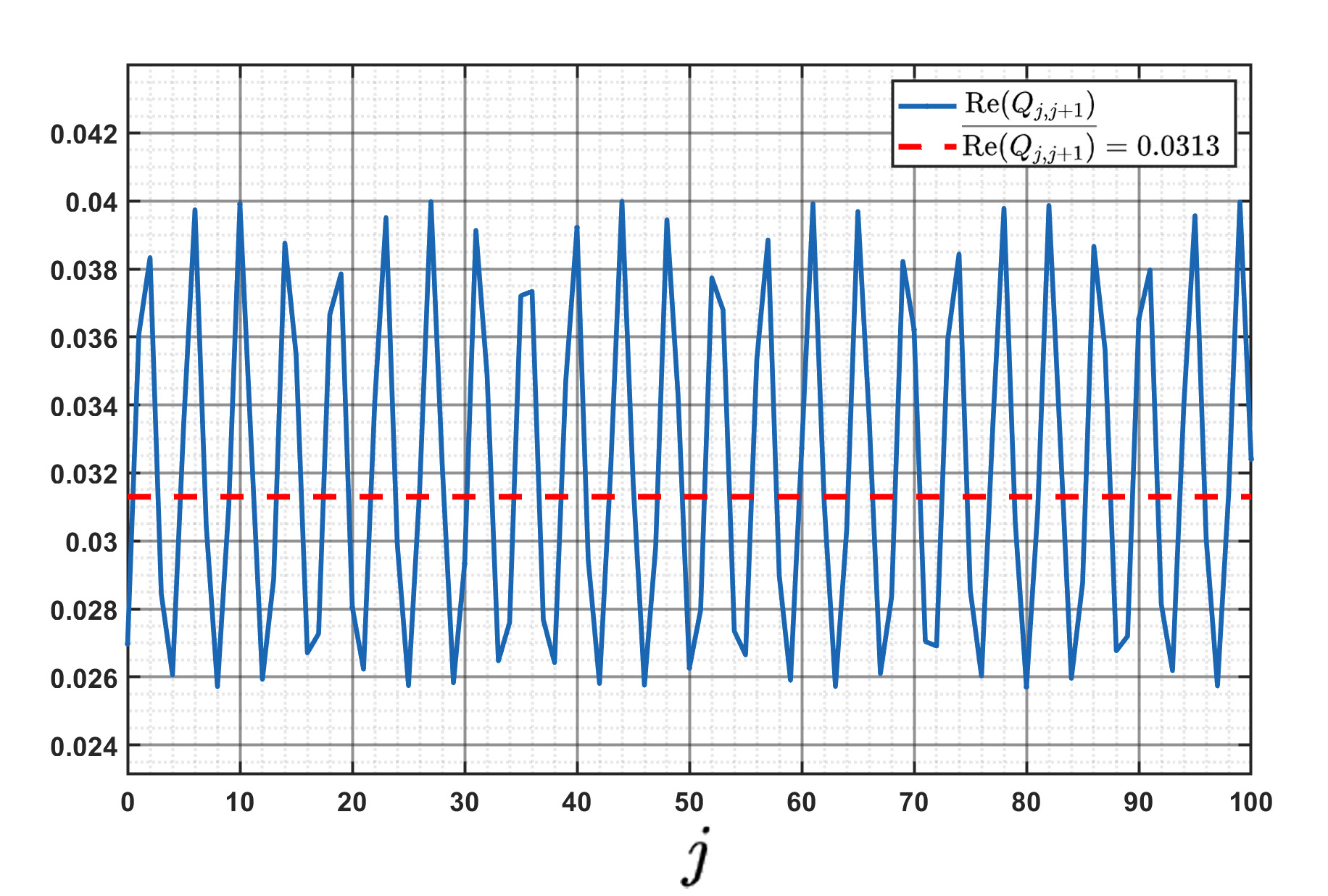}
    \caption{Schematic illustration of the spatial averaging of $\operatorname{Re}(Q_{j, j+1})$. The solid blue line denotes the spatial distribution of $\operatorname{Re}(Q_{j, j+1})$, while the dashed red line represents the corresponding spatial average, $\overline{\operatorname{Re}(Q_{j, j+1})}$. The noise parameters are $(W, \sigma) = (10, 5)$.}
    \label{fig:placeholder}
\end{figure}

The process of spatial averaging can be visually understood in Fig.~\ref{fig:placeholder}. This operation smooth the small spatial fluctuations of $\operatorname{Re}\left(Q_{j, j+1}\right)$, substituting the original distribution (blue solid line) with its average value (red dashed line).

Eventually, we have:
\begin{equation}\label{S60}
\begin{aligned}
\overline{\operatorname{Re}\left(Q_{j, j+1}\right)}&=\frac{\sigma^{2} \theta^{2}}{\sigma^{4}+2 W^{2} \theta^{4} \sin ^{2}(\pi \beta)}.
\end{aligned}
\end{equation}
 When $\sigma$ is non-zero, Eq. (\ref{S60}) reveals that $\overline{\operatorname{Re}\left(Q_{j, j+1}\right)}$ vanishes only when $W \to \infty$, confirming that noise can restore localized skin modes in quasi-periodic systems under dynamic evolution.

Interestingly, $\overline{\operatorname{Re}\left(Q_{j, j+1}\right)}$ exhibits non-monotonic dependence on noise intensity $\sigma$: it first increases then decreases with $\sigma$. The maximum of Eq. (\ref{S60}) occurs at $\sigma_{\max }=\left(2 W^{2} \theta^{4} \sin ^{2}(\pi \beta)\right)^{1 / 4}$.  This behavior qualitatively explains the previously observed non-monotonic relaxation time with respect to $\sigma$ and $W$.

Finally, we analyze the short-time behavior ($t\to 0$) of $\operatorname{Re} Q_{j, j+1}$:
\begin{equation}\label{short-time}
\begin{aligned}
\operatorname{Re} Q_{j, j+1}
&= \lim _{t \rightarrow 0} \int_{0}^{t} \cos [W(\cos (2 \pi \beta j)-\cos (2 \pi \beta(j+1)))(t-\tau)] e^{-\frac{\sigma^{2}}{\sigma^{2}}\left[(t-\tau)-\frac{1}{\theta}\left(1-e^{-\theta(t-\tau)}\right)\right]} d \tau \\
&= t-\frac{1}{6}\left[4W^2\sin^2(\pi\beta(2j+1))\sin^2(\pi\beta)+\frac{\sigma^2}{\theta}\right]t^3+\mathcal{O}(t^{4}).
\end{aligned}
\end{equation}

Eq. (\ref{short-time}) demonstrates that quasi-periodicity and noise only affect wavepacket dynamics at $\mathcal{O}(t^3)$ in the short-time limit. This explains why the spread moment and the average position curve coincide in the initial stages of evolution, as observed in the main text.

\section{Dynamic scaling analysis}
In this section, we will extend the previously derived master equation defined on lattice to a continuous form. Furthermore, we can use it to explain the scaling behavior of the wave packet evolution that we observe in the main text. 
\subsection{Using master equations under continuous limits to explain the evolution behavior of wave packets}
 Firstly, take the continuous limit, we have:
\label{S61} \label{S62}
\begin{align}
\left\langle P_{j}(t)\right\rangle &= \rho(x, t) a, \label{S61} \\
\left\langle P_{j \pm 1}\right\rangle &= \rho(x \pm a, t) a \approx a\left[\rho \pm a \frac{\partial \rho}{\partial x}+\frac{a^{2}}{2} \frac{\partial^{2} \rho}{\partial x^{2}}\right] \label{S62},
\end{align} 
where $\rho(x, t)$ represents the probability density of particles appearing at position $x$ at time $t$, and $a$ represents the lattice constant. Firstly, consider the scenario under the long-time limit, by substituting Eq.~(\ref{S61}) and Eq.~(\ref{S62}) into Eq.~(13) in main text, we have:
\begin{equation}\label{S64}
\frac{\partial \rho(x, t)}{\partial t}=S \rho(x, t)+v \frac{\partial \rho(x, t)}{\partial x}+D \frac{\partial^{2} \rho(x, t)}{\partial x^{2}},
\end{equation}
where
\begin{equation}\label{S67}
\begin{aligned}
S &= 8 \overline{\mathrm{Re}(Q_{j,j+1})}\, \Delta^2, \\
v &= 8 \overline{\mathrm{Re}(Q_{j,j+1})}\, J \Delta a, \\
D &= 2 \overline{\mathrm{Re}(Q_{j,j+1})}\, (J^2 + \Delta^2) a^2.
\end{aligned}
\end{equation}
$S$ is the source term in the equation, which is the source of non-conservation of the probability of wave function evolution. $v$ is the drift velocity of the wave packet, and $D$ is the diffusion coefficient. To perform Fourier transform on the above equation, we have:
\begin{equation}\label{S68}
\frac{\partial \tilde{\rho}(k, t)}{\partial t}=\left[S+i k v-D k^{2}\right] \tilde{\rho}(k, t) ,
\end{equation}
where 
\begin{equation}\label{S69}
\tilde{\rho}(k, t)=\int_{-\infty}^{\infty} e^{-i k x} \rho(x, t) d x, \quad \rho(x, t)=\frac{1}{2 \pi} \int_{-\infty}^{\infty} e^{i k x} \tilde{\rho}(k, t) d k.
\end{equation}
By setting the initial condition $\rho(x, 0)=\delta(x)$ ($\tilde{\rho}(k, 0)=1$), we can obtain the solution:
\begin{equation}\label{S70}
\tilde{\rho}(k, t)=\exp \left[S t-i k v t-D k^{2} t\right].
\end{equation}
Perform inverse Fourier transform on Eq. (\ref{S70}), and finally obtain:
\begin{equation}\label{solution1}
\rho(x, t)=\frac{e^{S t}}{\sqrt{4 \pi D t}} \exp \left[-\frac{(x-v t)^{2}}{4 D t}\right].
\end{equation}
Based on the above solution, we can calculate the average position and spread moment of the wave packet at any time. Specifically, we can directly bring Eq. (\ref{solution1}) into the definition formula of the average position, so we have:
\begin{equation}
X(t)=\frac{1}{P_{\text {tot }}(t)} \int x \rho(x, t) d x ,
\end{equation}
where
\begin{equation}\label{S71}
P_{\text {tot }}(t)=\int \rho(x, t) d x=e^{S t}.
\end{equation}
So we obtain:
\begin{equation}\label{S72}
\Delta X\left( t\right)=X(t)=\frac{1}{e^{S t}} \int x \cdot \frac{e^{S t}}{\sqrt{4 \pi D t}} \exp \left[-\frac{(x-v t)^{2}}{4 D t}\right] d x=v t .
\end{equation}
Using the same method, we can calculate $\Sigma(t)$:
\begin{equation}\label{S73}
\Sigma(t)=\left[\frac{1}{P_{t o t}(t)} \int x^{2} \rho(x, t) d x-\frac{1}{P_{t o t}(t)^{2}}\left(\int x \rho(x, t) d x\right)^{2}\right]^{\frac{1}{2}}\\
=\sqrt{v^{2} t^{2}+2 D t-(v t)^{2}}=\sqrt{2 D t} .
\end{equation}

Let's now consider the situation under the short-time limit, as deduced earlier, when $t$ approaches $0$, $\overline{\mathrm{Re}(Q_{j,j+1})}\ \sim t$. At this point, the coefficients are time-dependent. So we can have:
\begin{align}
S(t) &= 8\Delta^{2}t, \\
v(t) &= 8J\Delta a t ,\\
D(t) &= 2\left(J^{2}+\Delta^{2}\right)a^{2}t.
\end{align}
As before, by performing Fourier transform on both sides of the equation at this time, we have:
\begin{equation}
\frac{\partial \tilde{\rho}(k, t)}{\partial t}=\left[8 \Delta^{2} t-8 i k  J \Delta a t-2 k^{2} \left(J^{2}+\Delta^{2}\right) a^{2} t\right] \tilde{\rho}(k, t).
\end{equation}
Using the initial condition $\tilde{\rho}(k, 0)=1$ to solve the equation, we have:
\begin{equation}
\tilde{\rho}(k, t)=\exp \left(\frac{1}{2} (8 \Delta^{2}-8 i k J \Delta a-2 k^{2}\left(J^{2}+\Delta^{2}\right) a^{2}) t^{2}\right).
\end{equation}
To perform the inverse Fourier transform on the above equation, we obtain:
\begin{equation}\label{S82}
\rho(x, t)=\frac{e^{4  \Delta^{2} t^{2}}}{\sqrt{4 \pi \left(J^{2}+\Delta^{2}\right) a^{2} t^{2}}} \exp \left[-\frac{\left(x-4  J \Delta a t^{2}\right)^{2}}{4 \left(J^{2}+\Delta^{2}\right) a^{2} t^{2}}\right].
\end{equation}
By bringing Eq. (\ref{S82}) to the definitions of mean position and spread moment, we have:
\begin{align}
\Delta X\left( t\right) &= 4  J \Delta a t^{2},\\
\Sigma(t) &= \sqrt{2 \left(J^{2}+\Delta^{2}\right)} a t.
\end{align}
Therefore, we can see that as $\overline{\mathrm{Re}(Q_{j,j+1})}\ $ approaches a constant from proportional to $t$ over time, the evolution behavior of the wave packet will change from ballistic transport to diffusion behavior.

\section{Static spectrum and Bott index of non-reciprocal AAH model}
In this section, we present a detailed analysis of the noise-free scenario in the non-reciprocal AAH model introduced in the main text. By calculating the first-order non-Hermitian Bott index and endpoint weight, we systematically characterize the steady-state spectral properties of the system.
\subsection{Skin Corner Weight}
To quantify boundary localization under OBC, we employ the \emph{endpoint weight}~\cite{PhysRevB.111.155121}:
\begin{eqnarray}  
w_{\mathrm{CW}}^{(2)}\left(E_{n}\right) \equiv \sum_{\mathbf{r}} \sum_{i=1}^{2} (-1)^{i} \left|\psi_{n}(\mathbf{r})\right|^{4} e^{-\left|\mathbf{r}-\mathbf{r}_{i}\right| / \lambda}, \label{w_CW} 
\end{eqnarray}  
where $\mathbf{r}$ denotes lattice sites, $\psi_{n}(\mathbf{r})$ is the right eigenfunction of eigenvalue $E_{n}$, and $\mathbf{r}_{i}$ ($i=1,2$) are the two boundary positions in a one-dimensional chain.  
Here $\lambda$ is a localization scale chosen such that $\lambda \ll L$ for system size $L$. The sign of $w_{\mathrm{CW}}^{(2)}(E_{n})$ indicates the polarization of the skin mode: positive (negative) values correspond to right (left) boundary localization, while its magnitude quantifies localization strength.  

\subsection{Topological Characterization}
Since the NHSE manifests as a topological phenomenon, energy-resolved topological invariants provide a diagnostic tool. For the present model, the first-order multipole Bott index is suitable~\cite{PhysRevB.111.155121}:
\begin{equation}
P_{x}(E_{a}) = \frac{1}{2\pi i} \mathrm{Tr}\left[\log\left(\mathcal{P}_{x}^{A} \mathcal{P}_{x}^{B\dagger}\right)\right],
\end{equation}
with the dipole operator
\begin{equation}
\mathcal{P}_{x}^{S} = U_{S}^{\dagger} \Big[ \sum_{R, \alpha} \exp\left(i \tfrac{2 \pi x}{L_{x}}\right) |R, \alpha\rangle\langle R, \alpha| \Big] U_{S},
\end{equation}
where $U_{S}$ ($S=A,B$) are unitary matrices obtained from the singular value decomposition of $h=H-E_{a}$, i.e., $h=U_{A}\Sigma U_{B}^{\dagger}$ with $\Sigma$ the diagonal singular-value matrix. The basis state $|R, \alpha\rangle$ corresponds to orbital $\alpha$ in unit cell $R$.  

A nonvanishing $P_{x}(E_{a})$ in a region of the complex energy plane signals the NHSE for OBC eigenenergies within that region.  

\subsection{Static spectrum and Bott index of non-reciprocal AAH model}
\begin{figure*}
    \centering
    \includegraphics[width=1\linewidth]{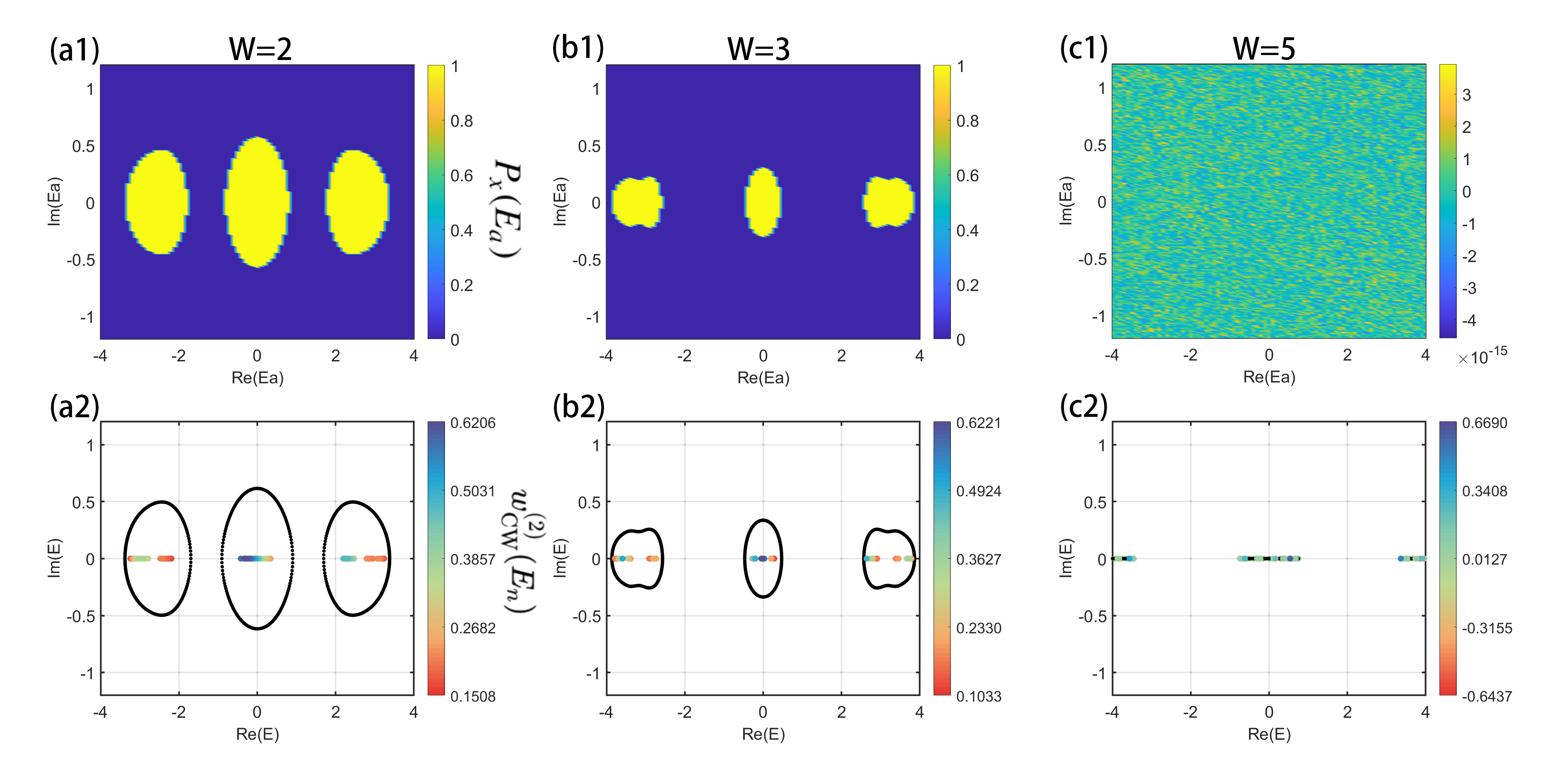}
    \caption{\label{fig1} (a1) (b1) (c1) show the distribution of $P_x(E_a)$ for the quasi-periodic Hatano-Nelson model under different $W$. 
    (a2) (b2) (c2) exhibit energy spectra under OBC (colored dots) and PBC (black dots) for various $W$. The color scale represents $w_{\mathrm{CW}}^{(2)}(E_{a})$ 
    of the eigenstates, with parameters $r_{1}=1$, $r_{2}=L$, and $\lambda=10$. Positive (negative) values of $w_{\mathrm{CW}}^{(2)}(E_{n})$ indicate localization of the skin effect modes at the right (left) boundary of the system. The model parameters are set to $J=\frac{3}{2} , \Delta=\frac{1}{2}$ with a lattice size of $L=100$. 
    }
    \label{fig1}
\end{figure*}
Fig.~\ref{fig1}(a)–(c) show the evolution of the Bott index $P_{x}(E_{a})$ across the complex energy plane, along with the spectra under periodic and open boundary conditions (PBC and OBC) as $W$ increases. The nonzero region of the Bott index corresponds to the loops formed by PBC spectra, which signifies the existence of the NHSE. Remarkably, the initially connected nonzero region of $P_{x}(E_{a})$ undergoes fragmentation into three distinct domains, which gradually shrink and ultimately vanish. Beyond the critical quasi-periodic strength ($W > W_{c}$), the region with nonzero Bott index disappears throughout the complex plane, consistent with the transition of all eigenstates into a localized phase and the disappearance of NHSE.

\section{Detailed analysis of Relaxation Times in the noise-induced dynamical skin effect}
\begin{figure*}
    \centering
    \includegraphics[width=1\linewidth]{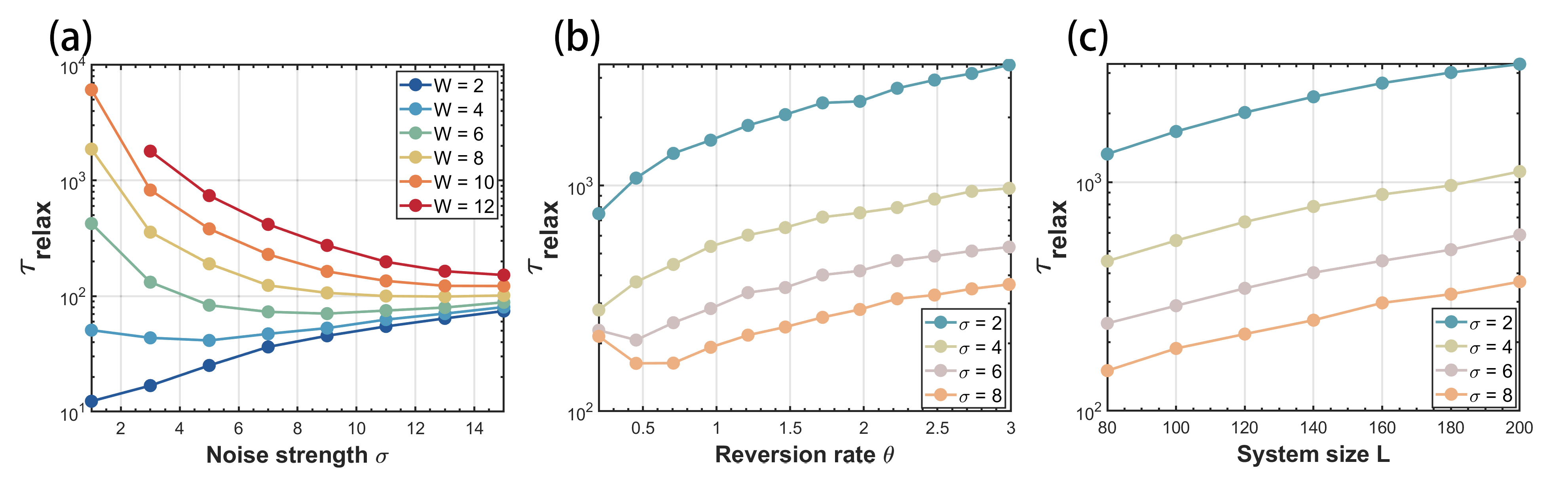}
    \caption{Relaxation time as a function of various model parameters. (a) Relaxation time $\tau_{\text{relax}}$ versus the OU noise strength $\sigma$ for different quasi-periodic potential strengths $W$. The system size is $L=100$, and the noise mean reversion rate is fixed at $\theta=1.0$. (b) Relaxation time $\tau_{\text{relax}}$ versus the noise mean reversion rate $\theta$ for different noise strengths $\sigma$. The quasi-periodic potential strength is $W=10$, and the system size is $L=100$. (c) Relaxation time $\tau_{\text{relax}}$ versus the system size $L$ for different noise strengths $\sigma$. The quasiperiodic potential strength is $W=10$, and the noise mean reversion rate is $\theta=1.0$. For all panels, the underlying non-reciprocal AAH model parameters are set to $J=3/2$ and $\Delta=1/2$. The initial state is chosen as a single-site wave packet localized at the center of the lattice. The relaxation time $\tau_{\text{relax}}$ is defined and extracted as the time required for the wave packet's absolute mean position $X(t)$ to reach $80\%$ of the system size $L$. All numerical data points are obtained by averaging over an ensemble of $100$ independent noise and disorder realizations.}
    \label{Relaxation_time}
\end{figure*}

To quantitatively characterize the relaxation dynamics of the noise-restored DSE, we define the relaxation time $\tau_{\text{relax}}$ as the time required for the mean position of an initially centrally localized wave packet to reach $0.8L$ during its evolution. Fig.~\ref{Relaxation_time} details its scaling behavior across various model parameters. As shown in Fig.~\ref{Relaxation_time}(a), for weak quasi-periodic potential ($W=2$), noise primarily disrupts the inherent coherent transport, causing $\tau_{\text{relax}}$ to increase monotonically. Conversely, in the strongly localized regime ($W=8, 10, 12$), noise supplies energy to overcome deep potential wells, sharply accelerating transport and reducing $\tau_{\text{relax}}$. 

Furthermore, we investigated the dependence of the relaxation time $\tau_{\text{relax}}$ on the mean reversion rate $\theta$ [Fig.~\ref{Relaxation_time}(b)].  As observed for a fixed quasiperiodic potential ($W = 10$), the evolution of $\tau_{\text{relax}}$ is highly sensitive to the underlying noise strength $\sigma$. Under relatively weak noise (e.g., $\sigma = 2, 4$), $\tau_{\text{relax}}$ increases monotonically with $\theta$. In contrast, under stronger noise (e.g., $\sigma = 6, 8$), the relaxation dynamics exhibit a non-monotonic trend.

Finally, the finite-size scaling of $\tau_{\text{relax}}$ with respect to the system size $L$ [Fig.~\ref{Relaxation_time}(c)] reveals that, across various noise parameters, $\tau_{\text{relax}}$ increases monotonically—and almost linearly—with the system size. This robust scaling confirms that the noise-restored DSE is indeed a novel physical phenomenon that persists in the thermodynamic limit, rather than a mere finite-size artifact.
\section{Further discussion on the drift velocity and diffusion coefficient}
\begin{figure*}
    \centering
\includegraphics[width=1\linewidth]{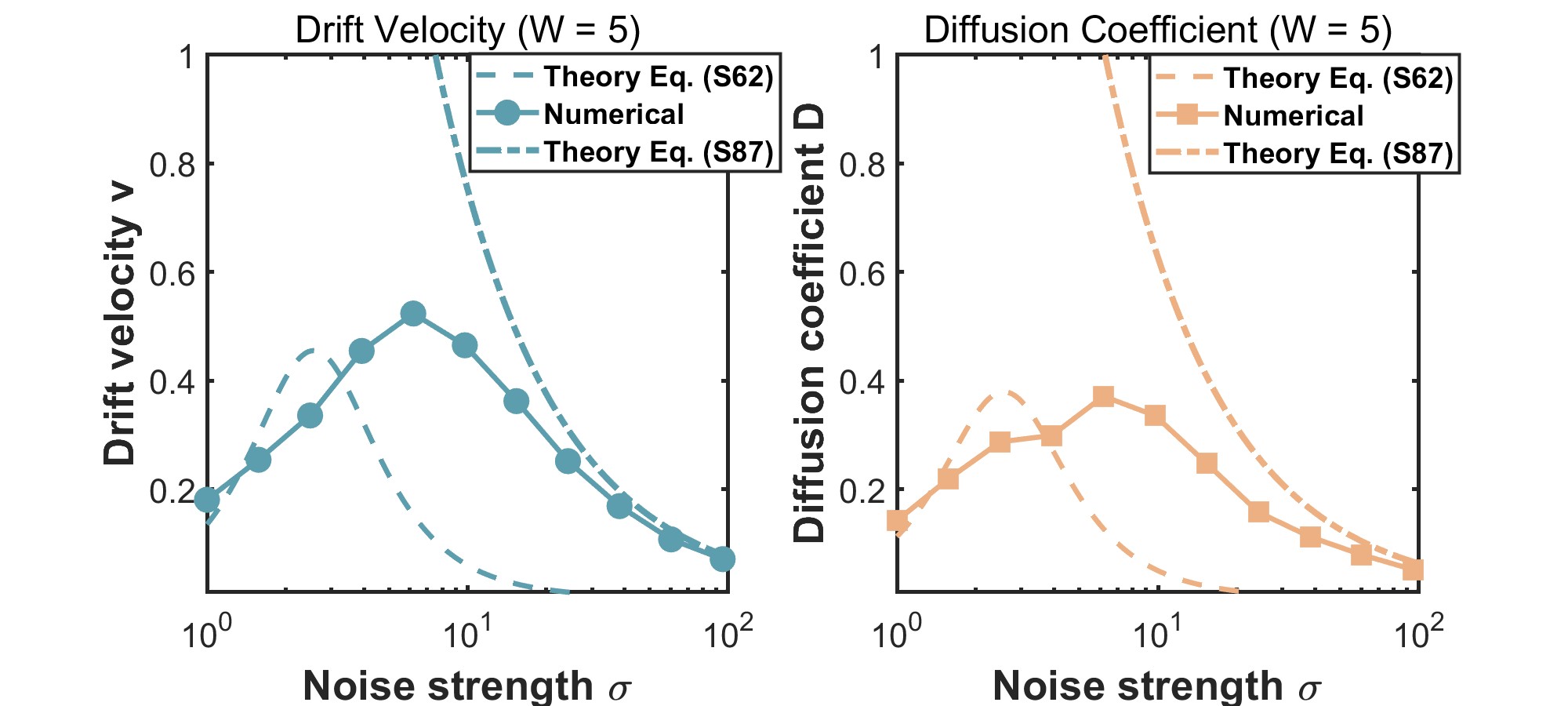}
    \caption{Numerical extraction and theoretical calculation of the drift velocity and diffusion coefficient under OU noise. The left and right panels display the drift velocity $v$ and the diffusion coefficient $D$, respectively, as functions of the noise strength $\sigma$. The numerical data (solid lines) are obtained by simulating the exact time evolution of a lattice system of size $L=100$. The initial state is set as a single-site excitation at the center of the lattice. For each $\sigma$, the time-dependent mean displacement $\langle X(t) \rangle$ and spatial variance $\langle \Sigma^2(t) \rangle$ are calculated and averaged over an ensemble of 100 independent OU noise realizations. To rigorously exclude the initial ballistic expansion and finite-size boundary reflection effects,  $v$ and $D$ are determined via linear fitting as $v = d\langle X(t)\rangle/dt$ and $D = \frac{1}{2} d\langle \Sigma^2(t)\rangle/dt$, respectively, within a dynamically chosen time window. This window corresponds strictly to the period when the  $\langle X(t) \rangle$ has traveled between $20\%$ and $70\%$ of the distance from its initial position to the right lattice boundary. The dashed lines represent the theoretical analytical predictions given by Eq. (\ref{S60}), which are derived using the long-time approximation for the phase accumulation integral. The dash-dotted lines represent the theoretical asymptotic values given by Eq. (\ref{large_sigma}), which are obtained by applying a short-time second-order Taylor expansion to the noise correlation decay. The static system parameters are fixed at $J=3/2$ , $\Delta=1/2$, $W=5$, and $\theta=1.0$.}
    \label{DV_W5}
\end{figure*}

In this section, we provide a detailed discussion comparing the numerical results and theoretical predictions for the drift velocity $v$ and diffusion coefficient $D$ of the wave packet across different parameter regimes. 
    As shown in Fig.~\ref{DV_W5}, the numerical curves (solid lines) for both $v$ and $D$ exhibit a non-monotonic behavior, initially increasing and then decreasing as the noise strength $\sigma$ increases. Comparing these numerical results with the theoretical predictions corresponding to Eq. (\ref{S60}) (dashed lines), we observe that in the intermediate noise regime (e.g., $\sigma \in [5, 10]$), Eq. (\ref{S60}) effectively captures this non-monotonicity, despite slight deviations near the maximum values. However, for substantially larger values of $\sigma$, the theoretical prediction of Eq. (\ref{S60}) consistently underestimates the numerical results.We attribute this discrepancy to the approximation made in the integral of Eq. (\ref{long_time}). Specifically, when $\sigma^2/\theta^2$ is an extremely large quantity, the integrand decays to zero rapidly for any $(t-\tau) > 0$ due to the exponential factor. Consequently, in the extreme strong-noise limit, the integral for $\text{Re} Q_{j, j+1}$ is overwhelmingly dominated by the ultra-short-time regime where $\tau \to 0$. 
    
\begin{figure*}
    \centering
    \includegraphics[width=0.6\linewidth]{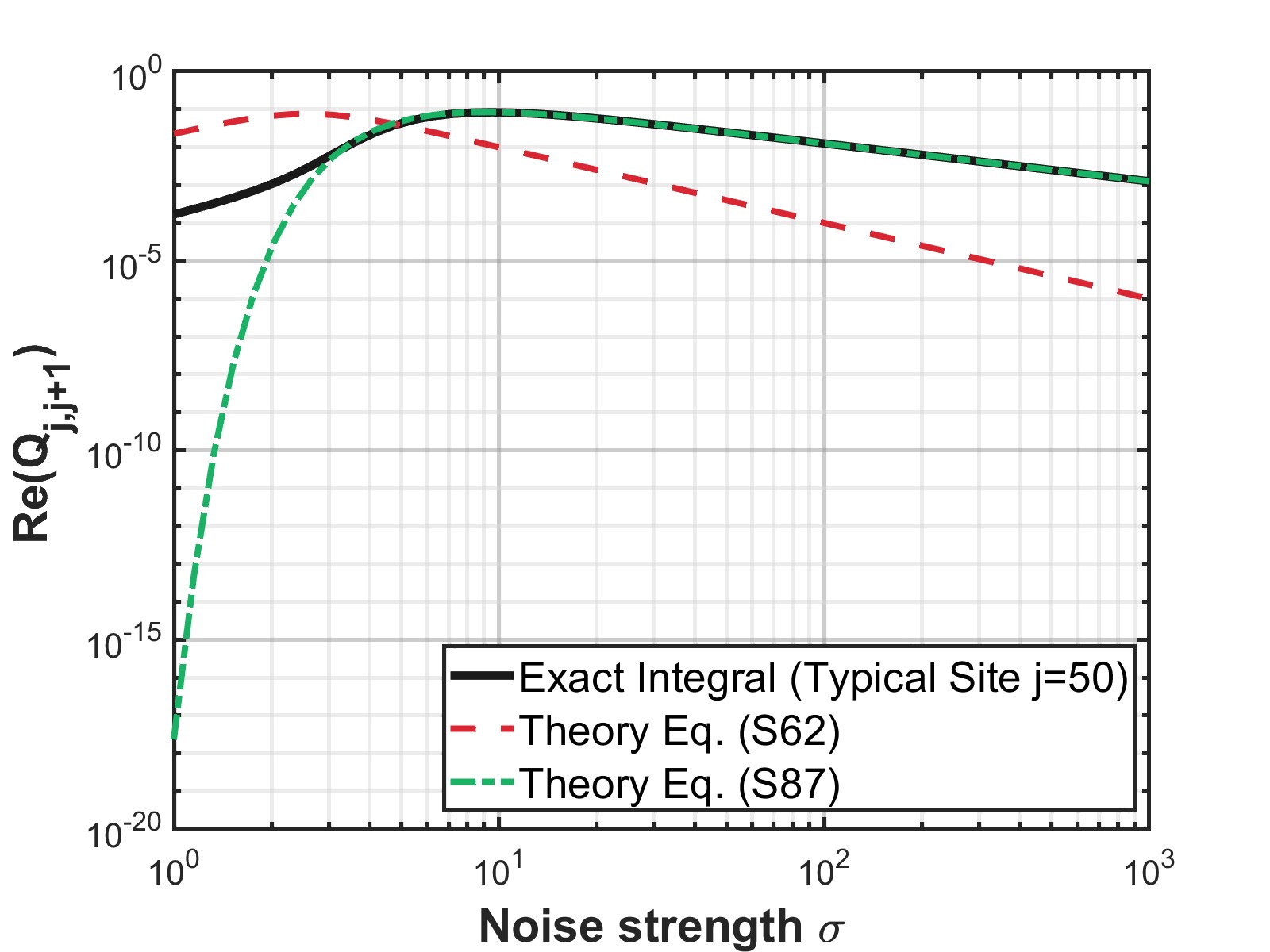}
    \caption{Exact numerical integration and analytical approximations for $\text{Re}(Q_{j,j+1})$. The black solid line represents the exact numerical evaluation of the unapproximated time-domain phase accumulation integral defining $\text{Re}(Q_{j,j+1})$. The red dashed line denotes the analytical expression provided by Eq. (\ref{S60}). The green dash-dotted line denotes the analytical asymptotic expression provided by Eq. (\ref{large_sigma}). Since the mathematical formulation of $\text{Re}(Q_{j,j+1})$ intrinsically depends on the local site index $j$ via the nearest-neighbor energy detuning $W[\cos(2\pi\beta j) - \cos(2\pi\beta(j+1))]$, a specific representative bulk lattice site $j=50$ is chosen to explicitly evaluate the exact numerical integral. Other parameters are $W=5$, $\theta=1.0$, and $\beta = (\sqrt{5}-1)/2$.}
    \label{compare}
\end{figure*}

    Specifically, as illustrated in Fig.~\ref{compare}, the exact numerical evaluation of the integral (black solid line) and the theoretical curve from Eq. (\ref{S60}) (red dashed line) agree well around $\sigma \approx 10$, but progressively diverge as $\sigma$ increases further.To establish a robust theoretical description in the strong-noise limit, we must re-evaluate the integral. Given that the integral is entirely dominated by the $\tau \to 0$ regime, it is necessary to perform a second-order Taylor expansion of the exponential term in the original function: $e^{-\theta\tau} \approx 1 - \theta\tau + \frac{1}{2}\theta^2\tau^2$. Substituting this expansion, we obtain:
\begin{equation}\label{large_sigma}
\begin{aligned}
\text{Re} Q_{j,j+1} &\approx \int_0^\infty \cos \big( W[\cos(2\pi\beta j) - \cos(2\pi\beta(j+1))] \tau \big) \exp\left( -\frac{\sigma^2}{\theta^2} \frac{1}{2}\theta\tau^2 \right) d\tau ,\\
&= \sqrt{\frac{\pi\theta}{2}} \frac{1}{\sigma} \exp\left( -\frac{\theta \big( W[\cos(2\pi\beta j) - \cos(2\pi\beta(j+1))] \big)^2}{2\sigma^2} \right),\\
&\approx \sqrt{\frac{\pi\theta}{2}} \frac{1}{\sigma}.
\end{aligned}
\end{equation}
The theoretical values predicted by Eq. (\ref{large_sigma}) are depicted by the green dashed line in Fig.~\ref{compare}. It is evident that this expression provides an excellent theoretical estimate in the extremely strong noise regime. Furthermore, the drift velocity $v$ and diffusion coefficient $D$ calculated using Eq. (\ref{large_sigma}), plotted as dash-dotted lines in Fig.~\ref{DV_W5}, demonstrate that Eq. (\ref{large_sigma}) successfully predicts the asymptotic behavior of the $v$ and $D$ curves under extremely strong noise.

\section{Initial state universality of the noise-restored dynamical skin effect}
\begin{figure*}
    \centering
    \includegraphics[width=1\linewidth]{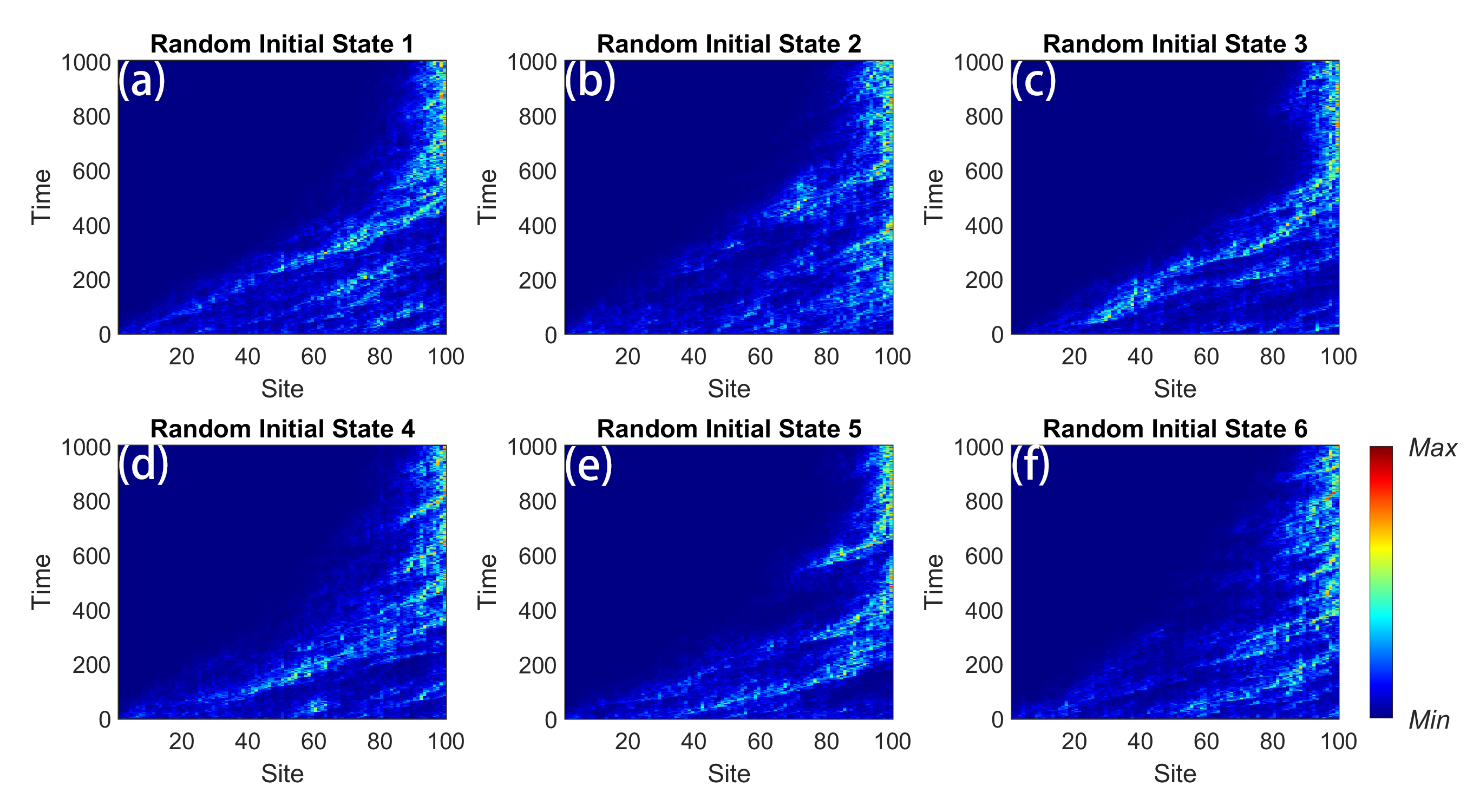}
    \caption{Initial state universality of the noise-restored dynamical skin effect. (a)-(f) Time evolution of the wave-function magnitude starting from six independent, completely random initial states. In each realization, both the amplitude and phase at every lattice site are uniformly randomized. The system parameters are fixed at $W=10$ and $\sigma=10$. Other model parameters are set to $J=3/2$ and $\Delta=1/2$, with a lattice size of $L=100$. Despite the highly disordered initial configurations, the noise persistently drives the probability density to accumulate at the right boundary. This confirms that the resurrection of the dynamical skin effect is independent of specific initial conditions.}
    \label{initial}
\end{figure*}

In the main text, our numerical simulations of the time evolution are consistently performed using an initial wave function strictly localized at the center of the lattice. To rigorously verify that the noise-restored DSE is a universal phenomenon rather than an artifact of this specific initial setup, we further investigate the system's dynamics starting from completely random initial states.As shown in Fig.~\ref{initial}(a)–(f), we simulate six independent realizations where both the amplitude and phase of the wave function at every lattice site are uniformly randomized. This constructs a globally extended and highly disordered initial configuration. The system parameters are fixed at $W=10$ and $\sigma=10$, corresponding to a regime deeply within the localized phase in the absence of noise. Strikingly, despite the extreme spatial disorder and the lack of initial localization, the environmental noise persistently breaks localization and drives the probability density to uniformly accumulate at the right boundary in all cases. These results unambiguously confirm that the resurrection of the DSE is highly robust and entirely independent of the choice of initial conditions.

\section{Noise-restored dynamical skin effect across diverse noise statistics}
\begin{figure*}
    \centering
    \includegraphics[width=1\linewidth]{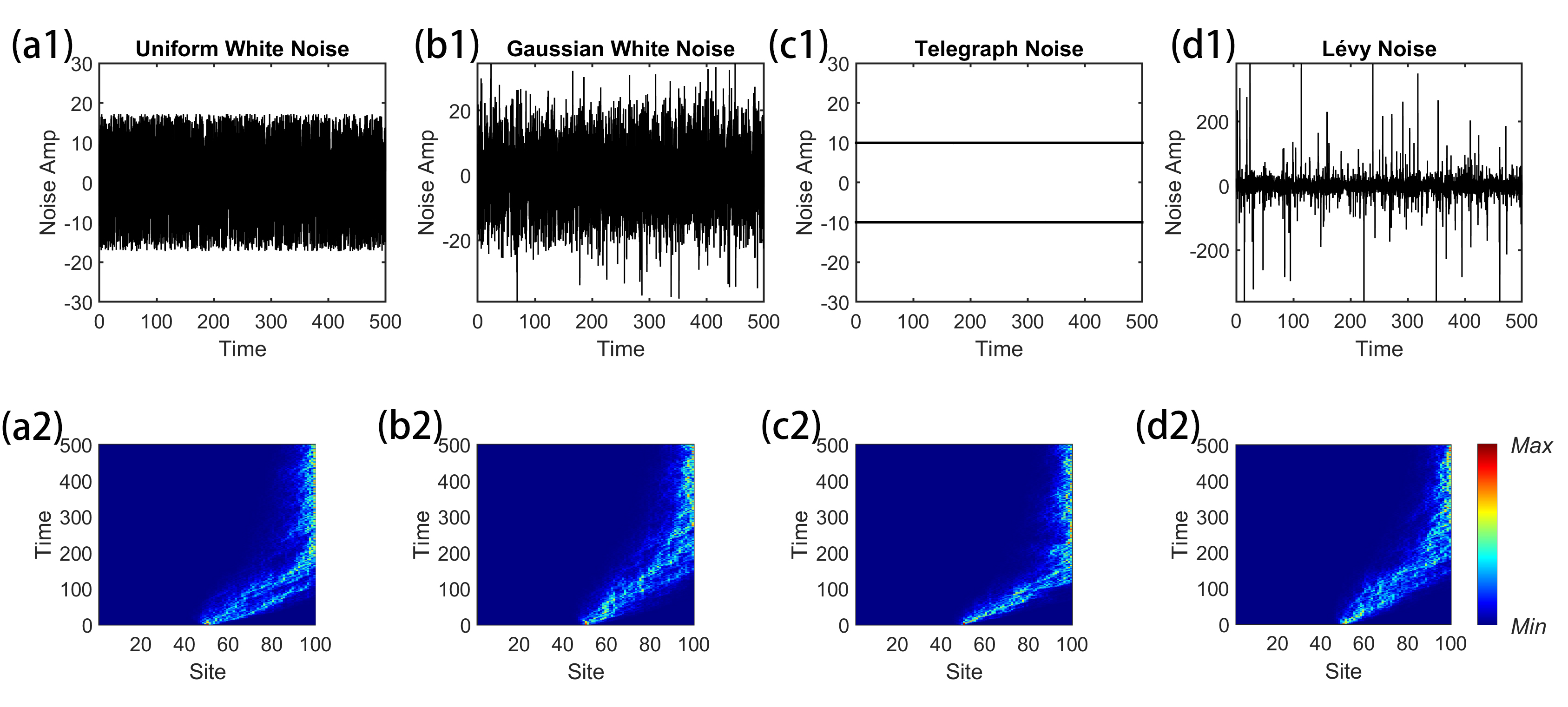}
    \caption{(a1)–(d1) Single realizations of the temporal noise profiles. (a2)–(d2) Time evolution of the wave-function dynamics for the quasi-periodic Hatano-Nelson model under the corresponding noise realizations shown in (a1)–(d1). The quasi-periodic potential strength is $W=10$. Other model parameters are set to $J=3/2$ and $\Delta=1/2$, with a lattice size of $L=100$. The initial state in (a2)–(d2) is chosen as a wave packet localized at the center of the lattice. The four distinct noise statistics are defined as follows: The uniform white noise consists of random variables uniformly distributed in the interval $[-10\sqrt{3}, 10\sqrt{3}]$. The Gaussian white noise is drawn from a normal distribution with zero mean and a standard deviation of $10$. The telegraph noise is a dichotomous random process switching between fixed amplitudes of $+10$ and $-10$, with $\gamma_{\text{flip}}=2$. The Lévy noise is generated from a symmetric $\alpha$-stable distribution with a stability index $\alpha = 1.5$, where an extreme-value truncation limit of $\pm 500$ is introduced in the numerical calculations to prevent computational divergence.}
     \label{Different_Noise}
\end{figure*}

To demonstrate the generality of the noise-restored DSE, we investigate the wave-packet dynamics under various types of temporal noise, as illustrated in Fig.~\ref{Different_Noise}. Specifically, we introduce uniform white noise, Gaussian white noise, telegraph noise, and Lévy noise into the quasi-periodic Hatano-Nelson model with a potential strength of $W=10$, which corresponds to a deeply localized regime in the absence of noise. 

The uniform and Gaussian white noises, as their names imply, consist of uncorrelated random variables drawn from uniform and normal distributions, respectively. Telegraph noise is a dichotomous random process that switches between two fixed amplitude states. Assuming an average flip rate of $\gamma_{\text{flip}}$, the waiting time before the next state inversion occurs is exponentially distributed. Consequently, the cumulative probability of a flip occurring within a time interval $t$ is given by $P(t) = 1-e^{-\gamma_{\text{flip}} t}$. For a sufficiently small numerical time step $\Delta t$, the probability of a state inversion occurring in each step can be reasonably approximated as $p_{\text{flip}} \approx \gamma_{\text{flip}} \Delta t$. In our numerical simulations, the Lévy noise is modeled as a sequence of independent and identically distributed symmetric $\alpha$-stable random variables\footnote{To be precise, the Lévy noise at each site is generated from the distribution $\xi \sim S(1.5, 0, 10, 0)$.}. We employ the standard Chambers-Mallows-Stuck algorithm \cite{chambers1976method,pogorzelec2023resetting} to rigorously generate this stochastic sequence. 

As clearly shown in Figs.~\ref{Different_Noise}(a2)--\ref{Different_Noise}(d2), despite the fundamental differences in the underlying noise statistics, the directed transport characteristic of the DSE is successfully restored in all cases. This confirms that the noise-induced resurrection of the DSE is a highly universal phenomenon.

\section{Noise resurrects dynamics skin effect in gain and loss systems }
The non-Hermitian skin effect in the previously discussed non-reciprocal AAH model arises from the asymmetry of the hopping amplitudes. In this section, we demonstrate that noise can still revive the skin effect caused by gain and loss.
Consider the model~\cite{PhysRevB.110.155144}:
\begin{equation}
\begin{aligned}
H= & \sum_{n=1}^{N} t_{1}\left(c_{n, a}^{\dagger} c_{n, b}+\text { H.c. }\right)+\sum_{n=1}^{N-1}\left(t_{2} c_{n, b}^{\dagger} c_{n+1, a}\right. 
 \left.+i t_{3} c_{n, a}^{\dagger} c_{n+1, a}+i t_{3} c_{n, b}^{\dagger} c_{n+1, b}+\text { H.c. }\right) \\
& +\sum_{n=1}^{N}\left(-i \Gamma c_{n, a}^{\dagger} c_{n, a}+i \Gamma c_{n, b}^{\dagger} c_{n, b}\right)  +\sum_{n=1}^{N}\left[\left( W \cos(2 \pi \beta n) + \xi_{n}(t) \right) \left( c_{n, a}^{\dagger}  c_{n, a}+ c_{n, b}^{\dagger} c_{n, b}\right)\right],
\label{loss_gain_H}
\end{aligned}
\end{equation}
where $c_{n, \alpha}$ $(c_{n, \alpha}^{\dagger})$ represents the annihilation (creation) operator of spinless fermions on the $\alpha$ sublattice of the $n$-th unit cell. $t_{1,2,3}$ and $\Gamma$ are are real parameters. $W$ represents quasi-periodicity strength, and $\beta$ is an irrational number. $\xi_{l}(t)$ is the OU noise in the $l$-th unit cell. The eigenstates of this model display a bipolar skin effect: in the OBC spectrum, a fraction of the eigenstates accumulate at the left edge, while the remaining portion localize at the right edge.

\begin{figure}  
    \centering
    \includegraphics[width=0.8\linewidth]{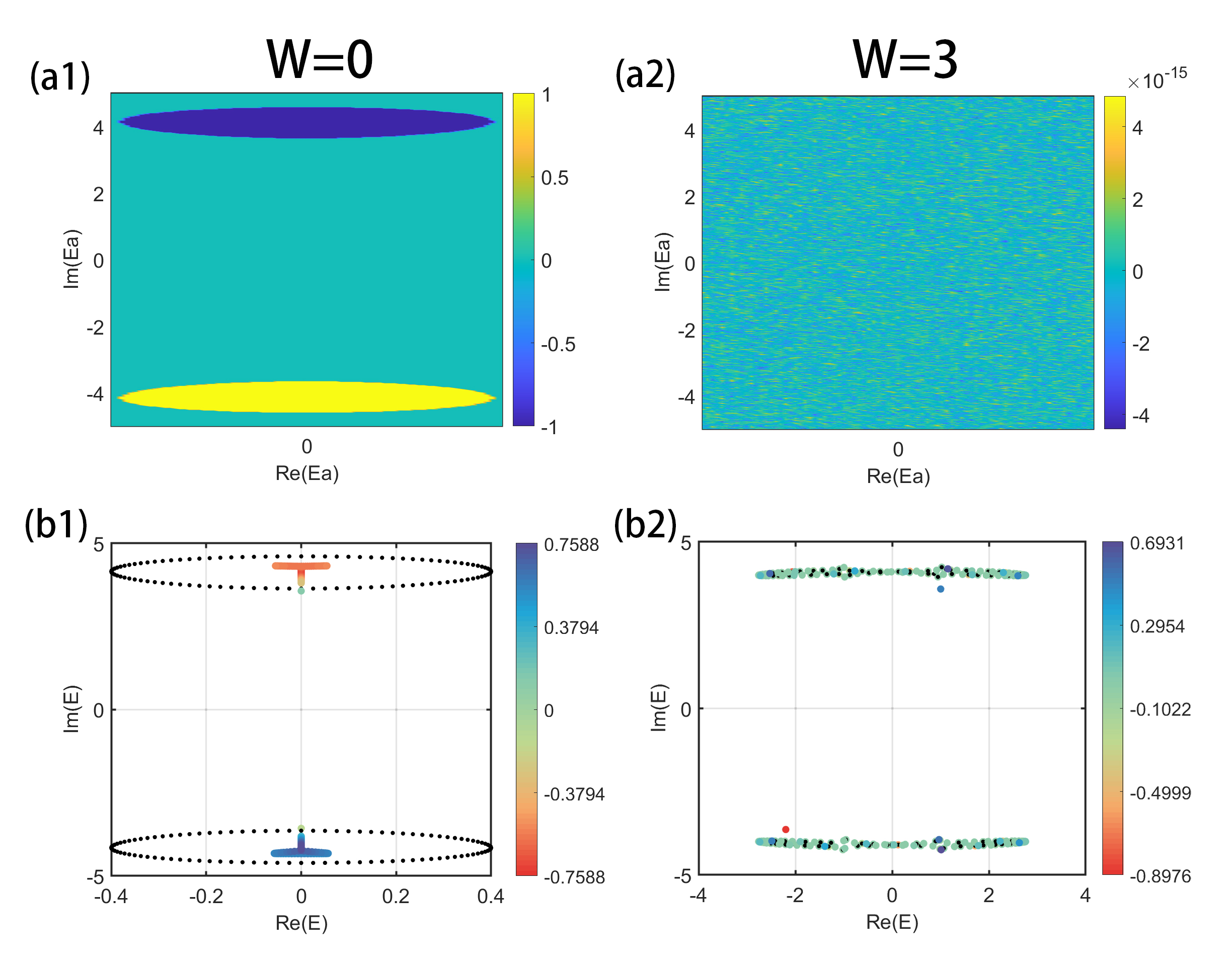}
    \caption{(a1) (a2) The first-order non-Hermitian Bott index for the model (\ref{loss_gain_H}) under different $W$. (b1) (b2) Energy spectra under OBC (colored dots) and PBC (black dots) for various $W$. The color scale represents $w_{\mathrm{CW}}^{(2)}(E_{a})$ of the eigenstates, with parameters $r_{1}=1$, $r_{2}=L$, and $\lambda=10$. Positive (negative) values of $w_{\mathrm{CW}}^{(2)}(E_{n})$ indicate localization of the skin effect (SE) modes at the right (left) boundary of the system. The model parameters are set to $\left(t_{1}, t_{2}, t_{3}, \Gamma\right)=(i, 2 i, 0.2,3.5)$ with a lattice size of $L=100$.}
    \label{figS1}
\end{figure}

Fig.~\ref{figS1} displays the energy spectrum and $P_{x}(E_{a})$ under varying $W$ for this model . For parameters $(t_{1}, t_{2}, t_{3}, \Gamma, W) = (i, 2i, 0.2, 3.5, 0)$, the energy spectrum bifurcates into two distinct segments. The eigenvalues in the upper half correspond to left skin states, while those in the lower half represent right skin states. Remarkably, when the quasi-periodic strength increases to $W=3$, the PBC spectrum no longer encloses a finite area, and the non-zero regions of $P_{x}(E_{a})$ vanishes completely. This transition implies that all eigenstates become localized, accompanied by the complete suppression of the NHSE.

\begin{figure*}
    \centering
    \includegraphics[width=1.05\linewidth]{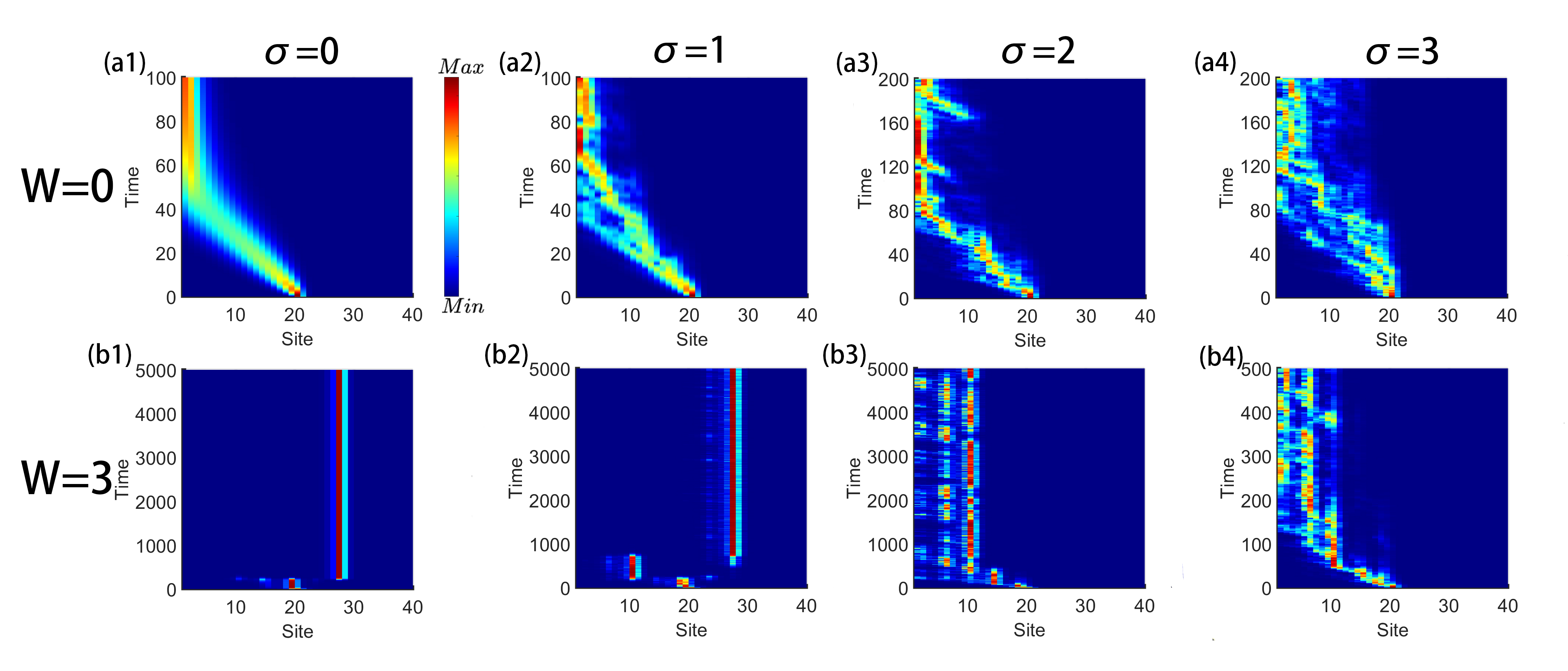}
    \caption{(a)(b)(c) The evolution of the wave function over time under noise strengths $\sigma=0$, $\sigma=1$ and $\sigma=3$. The model parameters are set to $\left(t_{1}, t_{2}, t_{3}, \Gamma,W\right)=(i, 2 i, 0.2,3.5,0)$ and $\left(t_{1}, t_{2}, t_{3}, \Gamma,W\right)=(i, 2 i, 0.2,3.5,3)$ with a lattice size of $L=40$. The initial state is selected as the state localized at the center of the one-dimensional chain.}
    \label{figS2}
\end{figure*}

The time evolution of the wave function under different noise amplitudes $\sigma$ and quasi-periodic potential strengths $W$ is illustrated in Fig.~\ref{figS2}. System parameters are set to $(t_1, t_2, t_3, \Gamma, W) = (i, 2i, 0.2, 3.5, 0)$ and $(t_1, t_2, t_3, \Gamma, W) = (i, 2i, 0.2, 3.5, 3)$ with a lattice size of $L = 40$. The initial state is a wave packet localized at the center of the chain.

 In the absence of a quasi-periodic potential ($W = 0$), the system exhibits inherent DSE due to NHSE, as shown in Fig.~\ref{figS2}(a). Introduction of OU noise ($\sigma = 1$), Fig.~\ref{figS2}(a2)-(a4) introduces mild fluctuation during propagation but does not influence the overall boundary accumulation. 

A finite quasi-periodic potential ($W=3$) induces localization, completely suppressing the DSE in noiseless case. In the absence of noise ($\sigma = 0$), as shown in Fig.~\ref{figS2}(b1), the wave function will jump slightly to the right and then remain localized at the position after the transition. This is a characteristic behavior of the wave function evolution in the gain loss system, which has been discussed in many previous articles \cite{z9m1-3mwb,weidemann2021coexistence}. Remarkably, introducing OU noise ($\sigma = 1,2,3$), as shown in Fig.~\ref{figS2}(b2)-(b4), resurrects undirectional transport and boundary accumulation gradually. DSE is gradually restored, and the wave packet will only stabilize at the boundary at the end of evolution after the noise intensity reaches a certain threshold.
When the noise intensity $\sigma$ reaches 3, The wave packet delocalizes, and eventually accumulates at the boundary, demonstrating a noise-induced revival of the DSE.

\bibliography{reference}